\documentclass[reprint, showpacs, amsmath, citeautoscript, amssymb, aps, prx, floatfix, longbibliography]{revtex4-2}
\pdfoutput=1

\usepackage{graphicx}
\usepackage{xcolor}
\usepackage{dcolumn}% Align table columns on decimal point
\usepackage{bm}% bold math
\usepackage{array}
\usepackage{hyperref}% add hypertext capabilities
\usepackage{siunitx}
\usepackage[capitalize]{cleveref}
\usepackage{multirow}
\usepackage{ulem}
\usepackage{tabularx, booktabs}
\usepackage{tensor}
\graphicspath{{figures/}}
\renewcommand\arraystretch{1.1}
\renewcommand{\vec}{\mathbf}
\newcommand{\tri}{\texorpdfstring{\textsubscript{3}}{3}}
\newcommand{\er}{ErTe\tri}
\newcommand{\tm}{TmTe\tri}
\newcommand{\R}{\textit{R}}
\newcommand{\rte}{\textit{R}Te\tri}
\newcommand{\pdx}{Pd\textsubscript{x}\rte{}}
\newcommand{\aax}{\textit{a}-axis}

\newcommand{\cax}{\textit{c}-axis}
\newcommand{\ti}{$T_{CDW1}$}
\newcommand{\tii}{$T_{CDW2}$}
\newcommand{\rms}{\textsubscript{rms}}

\newcommand{\qc}{$\vec{q_c}$}
\newcommand{\qa}{$\vec{q_a}$}
\newcommand{\degree}{$^\circ$}

\newcolumntype{Y}{>{\centering\arraybackslash}X}
\newcolumntype{C}[1]{>{\centering\let\newline\\\arraybackslash\hspace{0pt}}m{#1}}

\begin{document}
%\preprint{APS/123-QED}

\title{Evidence for realignment of the charge density wave state in ErTe\tri{} and TmTe\tri{} under uniaxial stress via elastocaloric and elastoresistivity measurements}

\author{J. A. W. Straquadine}
\email{jstraq@stanford.edu}
\affiliation{Geballe Laboratory for Advanced Materials and Department of Applied Physics, Stanford University, California 94305, USA}
\affiliation{Stanford Institute for Materials and Energy Sciences, SLAC National Accelerator Laboratory, 2575 Sand Hill Road, Menlo Park, CA 94025, USA}

\author{M. S Ikeda}
\affiliation{Geballe Laboratory for Advanced Materials and Department of Applied Physics, Stanford University, California 94305, USA}
\affiliation{Stanford Institute for Materials and Energy Sciences, SLAC National Accelerator Laboratory, 2575 Sand Hill Road, Menlo Park, CA 94025, USA}

\author{I. R. Fisher}
\affiliation{Geballe Laboratory for Advanced Materials and Department of Applied Physics, Stanford University, California 94305, USA}
\affiliation{Stanford Institute for Materials and Energy Sciences, SLAC National Accelerator Laboratory, 2575 Sand Hill Road, Menlo Park, CA 94025, USA}

\date{\today}

\begin{abstract}
	We report the evolution of the charge density wave (CDW) states in the quasi-2D rare-earth tritellurides (\rte{} for \R{}=Er, Tm) under the influence of in-plane uniaxial stress.
	Measurements of the elastocaloric effect, resistivity, and elastoresistivity allow us to demonstrate the importance of in-plane antisymmetric strain on the CDW and to establish a phase diagram.
	We show that modest tensile stress parallel to the in-plane \aax{} can reversibly switch the direction of the ordering wavevector between the two in-plane directions, and present a free energy expansion which reproduces the general structure of the observed phenomena.
	This work opens a new avenue in the study of \rte{} in its own right, and more generally establishes \rte{} as a promising model system for the study of strain-CDW interactions in a quasi-2D square lattice.
\end{abstract}

\maketitle

\section{Introduction}\label{sec:intro}

	% CDWS + SCES = COMPLICATED + NEED MODEL SYSTEMS
	The family of rare-earth tritelluride materials (\rte, \R=La-Nd, Sm, Gd-Tm, Y) has been studied for decades\cite{DiMasi1995,Gweon1998} as a model system for unidirectional charge density wave (CDW) formation in quasi-2D metals with an almost square lattice.
	Recent studies\cite{Johannes2008,Eiter2013,Maschek2018} have pointed out the inadequacy of the original picture of a nesting-driven\cite{Peierls1955,DiMasi1995,Gweon1998}, Peierls-like CDW transition.
	The simple and well-understood electronic structure, weak electronic correlations, and absence of strong magnetic fluctuations all make \rte{} a promising system to improve understanding of the origins and consequences of non-Peierls CDW ordered states.

	% STRAIN AND PRESSURE, %	\paragraph{strain technique motivations}
	While the mechanisms driving CDW formation can be material-dependent\cite{Chan1973,Requardt2002,Leroux2015,Maschek2016,Maschek2018}, there is an overarching consensus that strong coupling between the electronic and lattice degrees of freedom is crucial.\cite{Johannes2008}
	As such, modifying the lattice with hydrostatic pressure, chemical pressure, or uniaxial stress can produce substantial changes in the CDW state.\cite{Hamlin2009,Soumyanarayanan2013,Flicker2015,Zocco2015,Lei2020,Sharma2020}
	The response of the CDW to such perturbations contains a wealth of information about the host material, ordered state, and the phase transition.

	% Summary of results
	In this paper, we report investigations of the effect of in-plane uniaxial stress on unidirectional CDW states in \er{} and \tm{}.
	We provide evidence of in-plane realignment of the CDW wavevector under modest and practically-accessible stresses.
	It appears that the weak orthorhombicity acts as a bias ``field'' on the direction of the CDW, and that suitable uniaxial stress can overcome this bias and reverse the anisotropy of the CDW gap.
	Through this comprehensive study of the phases and phase transitions, we determine a stress-temperature phase diagram and establish \rte{} as a promising model system for the study of strain-CDW interactions in a quasi-2D square lattice.

	% MODEL SYSTEMS
	Unidirectional CDW states break both translational and rotational symmetries.
	The rotational component of the electronic order parameter will couple strongly to externally induced strains which break the same symmetry.
	However, the situation becomes less clear in the presence of disorder or competing instabilities,\cite{Nie2014,Fernandes2016,Nie2017,Fernandes2019} as in the case of CDW order in the cuprate superconductors \cite{Tabis2014a, Comin2015a, Gerber2015,Jang2016, Wen2019}.
	The interplay between the CDW and superconducting states in those materials is currently a subject of debate.
	To the extent that the CDW phenomena can be reproduced and studied in structurally similar, yet physically simpler materials, one can endeavor to improve understanding of the role played by charge order by means of analogy.
	\rte{} is regarded as a model system for unidirectional CDW formation, but the evolution of the CDW under in-plane, uniaxial stress has not been evaluated.
	This article opens a previously unexplored avenue of study into CDW formation in \rte{} by introducing uniaxial stress as a tuning parameter.

	% why do?
	We employ two uniaxial-stress techniques, namely the elastocaloric effect and the elastoresistivity, to elucidate the phase diagram of \rte{}.
	The elastocaloric effect (ECE) is a thermodynamic probe which measures the effects of strain on the entropy landscape.
	ECE measurements mimic the singular behavior at a phase transition observed in heat capacity $C_p$, but the selective sensitivity of ECE only to strain-dependent degrees of freedom leads to a much lower background signal at high temperatures than $C_p$.\cite{Hristov2019}
	In contrast, elastoresistivity (ER) is a transport probe that is sensitive to strain-induced changes in the Fermi surface topology, density of states at the Fermi surface, and scattering processes such as critical fluctuations.

\begin{figure}
		\centering
		\includegraphics[width=\columnwidth]{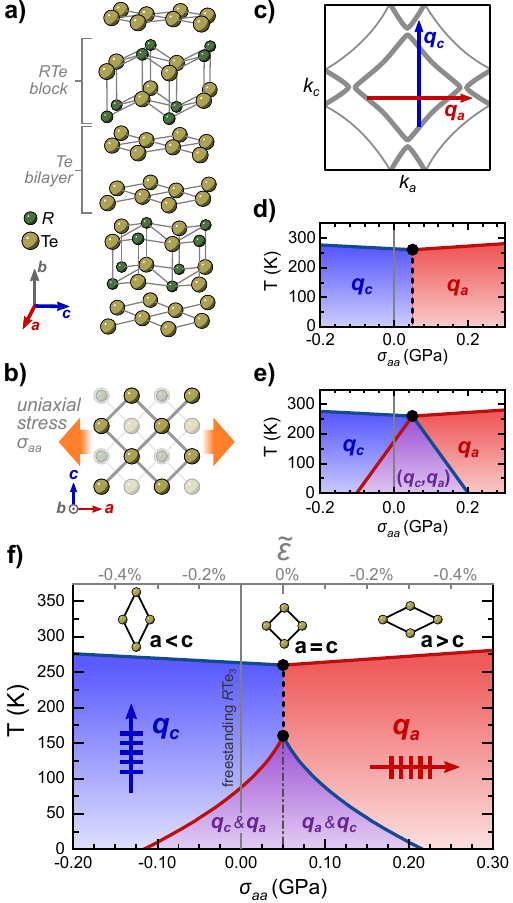}
		\caption{
			(a) Crystal structure of \rte{}.
			The stacking of the \R Te block layers produces a glide plane which breaks $C_4$ symmetry.
			(b) Crystal structure viewed along the out-of-plane $b$-axis.
			Tensile stress along the \aax{} can invert the native orthorhombic distortion, $a\approx0.999c$.
			(c) Schematic of the Fermi surface of \rte{}, neglecting the small $b$-axis dispersion and bilayer splitting.
			(d) Temperature-uniaxial stress phase diagram described by \cref{eq:fe1} for $g>0$.
			Solid lines indicate continuous phase transitions, and dashed lines indicate first order transitions.
			The stress axis is defined relative to a free-standing orthorhombic \rte{} crystal.
			(e) Phase diagram of \cref{eq:fe1} for $g<0$.
			A region of coexisting CDW order with both wavevectors opens at low temperatures, and all transitions are continuous.
			(f) Phase diagram of \cref{eq:fe2} incorporating higher order terms, which preserves the first order transition (between multicritical points) and the coexistence region.
			While missing from this model, resistivity measurements (\cref{sec:resistivity}) suggest the first order transition persists to lower temperature as well (dash-dot line).
			Coefficients appropriate for \er{} are used in calculating (d)-(f).
		}
		\label{fig:glrte3}
	\end{figure}

	%% TRITELLURIDES: structure
	Members of the rare-earth tritelluride family consist of bilayers of square Te nets separated by a buckled rock-salt layer of \R Te, shown in \cref{fig:glrte3}(a).
	The crystal structure of \rte{} belongs to the orthorhombic space group \textit{Cmcm}, and the standard crystallographic setting defines the $b$-axis normal to the planes.
	The orthorhombicity arises due to a glide plane along the in-plane \cax{}, which dictates the stacking of the \R Te slab layers and generates a slight difference in the two in-plane lattice parameters: $a\approx 0.999c$, hence this material comprises an ``almost square'' lattice.
	For all \R{}, the material undergoes a transition to a unidirectional, incommensurate CDW phase with the wavevector $q_c \approx (0,0,2c^*/7)$, where $c^*$ is the reciprocal lattice spacing.
	Chemical pressure tunes the transition temperature from above 500~K in LaTe\tri{} to 250~K in \tm{}.
	A second incommensurate CDW perpendicular to the first, with wavevector $q_a \approx (5a^*/7, 0, 0)$, appears for \R{}=Tb, Dy, Ho, Er, Tm at temperatures ranging from 41~K in TbTe\tri{}\cite{Banerjee2013} to 180~K in \tm{}.
	Throughout this work, we denote the higher transition temperature as \ti{} and the lower transition as \tii{}.
	In freestanding samples, \ti{} corresponds to a wavevector parallel to the \cax{}.

	% TABLE OF CONTENTS
	We begin in \cref{sec:landau} by developing a phenomenological, symmetry-motivated free-energy expansion and exploring the resulting phase diagrams.
	\Cref{sec:methods} covers the experimental methods, and \cref{sec:results} describe the results of elastocaloric effect, resistivity, and elastoresistivity measurements.
	We conclude in \cref{sec:disco} with a discussion of the broader impacts and future research opportunities inspired by this work.

	% Overview figure describing the basic concept

\section{Unidirectional CDWs in a quasi-tetragonal system}\label{sec:landau}
	Before describing our results in this specific model system, it is instructive to consider the general expectations for the free energy and the phase diagram.
	Consider a two-dimensional system with tetragonal symmetry which supports incommensurate unidirectional CDW order along both the $a$- and $c$-axes, where we keep the notation of $a$ and $c$ as in-plane lattice parameters for immediate comparison to \rte{}.

	Following the seminal work of McMillan\cite{McMillan1975}, but using the crystallographic coordinates of \rte{}, we take as our order parameters
		\begin{align}
			\psi_{j}(\vec{r}) = \psi_{j0}(\vec{r}) e^{i\vec{q_{j}}\cdot\vec{r}}
		\end{align}
	where ${j}=a,c$, and the subscript $0$ is used to indicate the absence of applied stress.

	Knowledge of the \rte{} family allows us to make several simplifying assumptions.
	The absence of commensurability effects (suggested by the smooth variation of \qc{} and \qa{} with temperature\cite{Malliakas2006,Ru2008b,Banerjee2013}) coupled with the low levels of CDW-pinning disorder (as observed in	x-ray diffraction\cite{Ru2008b,Banerjee2013}, STM\cite{Fang2007,Fu2016,Ralevic2016},	electrical transport\cite{Ru2006,Sinchenko2014} and quantum oscillations\cite{Ru2008c,Grigoriev2016a,Lei2020}) allows us to treat the wavevectors \qa{} and \qc{} as fixed, spatially uniform parameters.
	Also, as the two wavevectors are not parallel, the phases of the order parameters $\psi_a(\vec{r})$ and $\psi_c(\vec{r})$ are mutually independent.
	These assumptions together allow us to consider only the CDW gap magnitudes and suppress any gradient terms in our free energy expansion.
	For brevity, from this point forward we use the notation $\psi_a=|\psi_{a}|$, $\psi_c=|\psi_{c}|$.

	To introduce stress and strain terms into our free energy, we must also make a distinction between this idealized tetragonal model and the realistic orthorhombic structure of \rte{}.
	The stress and strain tensors in these two cases can be captured by a weakly temperature-dependent offset: $\varepsilon_{ij} + \varepsilon_{ij}^0 = \tilde{\varepsilon}_{ij}$, $\sigma_{ij} + \sigma_{ij}^0 = \tilde{\sigma}_{ij}$ where a tilde denotes the tetragonal case.\footnote{This assumption is not \textit{a priori} obvious.  While the difference in the \textit{a} and \textit{c} lattice parameters is small, and can in principle be compensated for by an externally induced strain of equal magnitude, the material would nevertheless remain fundamentally orthorhombic due to the glide plane (a non-symmorphic symmetry element that cannot be changed or removed solely by strain).  The fact that our observations reveal a rotation of the CDW wavevector indicates that the glide plane is not crucial in establishing the direction of the CDW wavevector.}

	Furthermore, we choose a basis for the stress, strain, and elastic constant tensors motivated by tetragonal symmetry.
	We use subscripts $A$ and $S$ for antisymmetric and symmetric in-plane components, respectively: $\tilde{\varepsilon}_A=(\tilde{\varepsilon}_{cc}-\tilde{\varepsilon}_{aa})/2$, $\tilde{\varepsilon}_S=(\tilde{\varepsilon}_{cc}+\tilde{\varepsilon}_{aa})/2$, and similar for the stresses $\tilde{\sigma}_A$ and $\tilde{\sigma}_S$.
	\footnote{Analogous stress and strain terms can be defined in the orthorhombic reference frame as well (e.g. $\varepsilon_A=(\varepsilon_{cc}-\varepsilon_{aa})/2$), however it should be noted that the subscripts $A$ and $S$ do not formally correspond to antisymmetric and symmetric strains; in the absence of tetragonal symmetry, these are no longer independent basis functions for different irreducible representations of the point group.}
	In this basis and neglecting out-of-plane contributions, the symmetrized and antisymmetrized elastic constants are defined as $\tilde{C}_S \approx C_S = 2(C_{aaaa}+C_{aacc})$ and $\tilde{C}_A \approx C_A=2(C_{aaaa}-C_{aacc})$, where the factor of two is added for convenience.

	Taking the tetragonal case as a reference, consider an expansion to fourth order in $\psi_a$ and $\psi_c$ of the Gibbs free energy (constant stress, constant temperature) given by:
	\begin{align}\label{eq:fe1}
		G_4 				&= G_\psi + G_\varepsilon + G_c\\
		G_\psi 			&= \frac{a_0t}{2}(\psi_a^2 + \psi_c^2) + \frac{b}{4}(\psi_a^2 + \psi_c^2)^2 + \frac{g}{2}\psi_a^2\psi_c^2\\
		G_\varepsilon	&= \frac{\tilde{C}_A}{2}\tilde{\varepsilon}_A^2 +  \frac{\tilde{C}_S}{2}\tilde{\varepsilon}_S^2 -2(\tilde{\sigma}_A\tilde{\varepsilon}_A + \tilde{\sigma}_S\tilde{\varepsilon}_S)\\
		G_c &=  \lambda \tilde{\varepsilon}_A(\psi_c^2 - \psi_a^2) + \eta\tilde{\varepsilon}_S(\psi_c^2 + \psi_a^2)\label{eq:gcoupling}
	\end{align}
	where $t=(T-T_c^0)/T_c^0$ is the reduced temperature and $T_c^0$ is the critical temperature in the hypothetical tetragonal crystal.
	Translational symmetry prevents the existence of bilinear terms involving the order parameters; the $\lambda$ and $\eta$ terms are the lowest order strain couplings allowed.
	The coefficients $a_0$, $b$ as well as the elastic constants $C_S$ and $C_A$ must be positive for stability.
	Empirically, the antisymmetric coupling constant $\lambda$ must be negative to stabilize the CDW parallel to the longer in-plane \cax{}.
	This is also supported by measurements of the thermal expansion below \ti{}\cite{Ru2008b}.
	Hydrostatic pressure experiments\cite{Hamlin2009,Zocco2015,Lei2020} suggest that the symmetric coupling coefficient $\eta$ must also be negative.
	The orthorhombic crystal structure implies that the lowest-order couplings between the order parameters and shear strains $\varepsilon_{ab}$, $\varepsilon_{bc}$, and $\varepsilon_{ac}$ are biquadratic and are neglected here.

	% Prior work, context
	Previous work has explored the \rte{} phase diagram with the assumption of $C_4$ symmetry without strain coupling\cite{Yao2006}, equivalent to the case $\lambda=\eta=0$.
	In this case, $g$ must be positive in order for a unidirectional CDW to form rather than a checkerboard state.
	In this case, finite strain coupling ``selects`` the CDW wavevector between the \qc{} and \qa{} states, separated by a first-order transition as shown in \cref{fig:glrte3}(d).
	In such a model, no second CDW transition is observed at lower temperatures.
	This would appropriately describe the phase diagram of \rte{} for \R{}=La, Ce, Pr, Nd, Sm, and Gd.

	The case for $g<0$, considered in \cref{fig:glrte3}(e), exhibits similar wavevector switching behavior but with a region of coexisting \qc{} and \qa{} states, bounded by a pair of second order transitions.
	At first glance, a vertical cut of this phase diagram for finite $\tilde{\sigma_A}$ appears to reproduce the cascade of phase transitions observed in \rte{} for \R{}=Tb, Dy, Ho, Er, and Tm.

	The measurements reported in \cref{sec:ece,sec:resistivity,sec:er} all suggest, however, that the real phase diagram exhibits both a first-order transition and a coexistence region; therefore neither of these two phase diagrams is sufficient.
	The phase diagram for the heavy rare earth compounds therefore requires the inclusion of higher order terms.
	One such model has been explored to eighth order for the case of \rte{} in the presence of disorder\cite{Fang2019}.
	In the pristine case, we have found that adding the two possible sixth order terms \cref{eq:fe1} suffices to capture the basic phenomenology of our observations:
	\begin{equation} \label{eq:fe2}
		G = G_4 + r(\psi_c^2+\psi_a^2)^3 + \gamma(\psi_c^4\psi_a^2 +\psi_c^2\psi_a^4)
	\end{equation}
	where $g>0$, $r>0$, and $\gamma\approx -r/2$.
	This phase diagram is shown in \cref{fig:glrte3}(f).
	Comparing our results to the physically distinct cases presented in \cref{fig:glrte3}(d-f) will provide insight into the nature of the CDW states and their mutual interactions.
	The experimental results described through the rest of this article will demonstrate that the stress-temperature phase diagram of \rte{} incorporates a first order transition as well as two independent CDW transitions, and therefore matches the structure of \cref{fig:glrte3}(f) more closely than \cref{fig:glrte3}(d) or (e).

\section{Experimental Methods}\label{sec:methods}

	\begin{figure*}
		\centering
		\includegraphics[width=\textwidth]{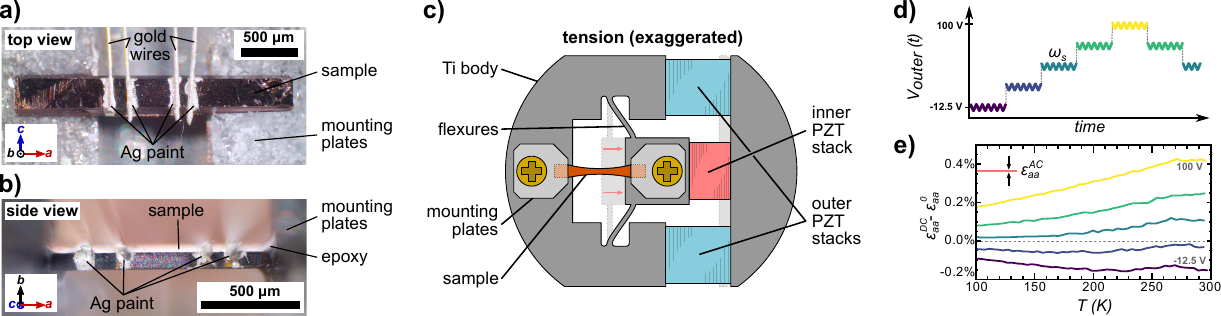}
		\caption{
			Description of the experimental setup.
			(a) Microscope photograph of the top and (b) side of a representative \er{} sample mounted in the stress cell, and contacted for \aax{} resistivity and elastoresistivity measurements.
			(c) Cartoon drawing of the CS-100 stress cell from \textit{Razorbill Instruments}.
			The outer two piezoelectric actuators were elongated ($V_\text{outer}>0$) and the inner is compressed ($V_\text{inner}<0$), resulting in tensile stress applied to the sample.
			(d) Schematic of the voltage applied to the outer stacks as a function of time.
			The actuator voltage was the sum of a stepwise DC offset (-25 to +100~V) with a small sinusoidal oscillation (ca. 5~V\rms, 30-80~Hz).
			Data was taken continuously while the sample temperature was ramped slowly (approx. 1~K/min), and the duration of each DC step was approximately 6.5 seconds.
			(e) DC component of the strain measured for a representative \er{} sample as a function of temperature and offset voltage $V_\text{outer}$.
			The temperature dependent response of the piezoelectric actuators contributes to some drift in strain for any given temperature.
			The height of the shaded bar in the upper left corner of (e) represents the maximum amplitude of the superimposed AC strain oscillation.
		}
		\label{fig:experiment}
	\end{figure*}
	Single crystals of \er{} and \tm{} are grown by a self-flux method described elsewhere.\cite{Ru2006}
	Further details of sample mounting, contacts, and thermometry can be found in \cref{app:methods}.

%	\paragraph{Stress, Razorbill}
	Uniaxial stress was generated using a commercially available stress cell, specifically the CS-100 from {\textit{Razorbill Instruments}}, in which a bar-shaped sample was suspended between a pair of mounting plates.
	The sample was then stressed by changing the spacing of the mounting plates \textit{in situ} by applying voltages to three piezoelectric actuators (stacks of lead zirconate titanate, or PZT) shown schematically in \cref{fig:experiment}.
	Two independent control voltages, $V_\text{inner}$ and $V_\text{outer}$  were applied to the inner and outer PZT stacks, respectively.
	$V_\text{outer}>0$ induced tensile stress in the sample and $V_\text{inner}>0$ induced compressive stress.

%	\paragraph{Control voltage waveforms}
	Measurements of ECE and ER require the superposition of both static and oscillating stresses.
	Signals of the form $V_\text{outer} = V_{dc} + V_{ac} \cos(2\pi f_s t)$ were used to drive the outer PZT stacks.
	$V_{dc}$ varied between \SI{-25}{V} and \SI{100}{V}, $f_s$ varies between \SI{20}{Hz} and \SI{90}{Hz}, and an oscillation amplitude of 5~V\rms{} was used for all measurements.
	The inner stack was compressed ($V_\text{inner}<0$) to keep the sample under tension and avoid buckling or delaminating these soft, layered samples.
	While we do not have a direct \textit{in situ} monitor for sample buckling during the experiment, bending and even partial delamination of some samples was observed at the conclusion of measurements with compressive stresses.

%	\paragraph{History dependence}
	In order to disentangle effects which rely on the temperature and strain history of the sample, data were collected using two different protocols.
	The first was to slowly sweep temperature while more rapidly stepping $V_\text{outer}$ up and down.
	On each step, the DC component was changed by $\pm$12.5~V.
	Unless otherwise noted, data presented in the figures only shows data taken for steps which were increasing in voltage.

	The second protocol used was to sweep temperature up and down for a fixed value of $V_{dc}$.
	This is not equivalent to measurement at constant strain.
	While the strain was monotonic in $V_{dc}$ at any given temperature, temperature dependence of the displacement-per-volt in the PZT actuators caused the actual strain to drift as a function of temperature, as shown in \cref{fig:experiment}(e).

%	\paragraph{Strain Measurement}
	The displacement $\Delta L$ of the jaws was measured using the capacitive sensor built into the CS-100 cell along with a bridge circuit described in \cref{app:capbridge}.
	Thermal expansion mismatch between the sample and the mounting materials, as well as the finite stiffness of the epoxy and cell itself, both produce temperature-dependent offsets of stress.
	These effects are discussed at length in \cref{app:nonideal}.
	The arguments of this paper, however, rely only on relative changes of stress--not the absolute stress values.

%	\paragraph{Synthesis/prep/mounting}
	Samples are cleaved and cut by hand with a scalpel into rectangular bars of \SIrange{1.3}{2.4}{mm} in length, \SIrange{250}{500}{\micro m} in width, and \SIrange{25}{130}{\micro m} in thickness.	

	The quantity measured by the AC elastoresistivity technique in principle consists of a real (in-phase) component which contains information about the strain derivative of the resistivity, and an imaginary (in-quadrature) component which contains information about dynamical or hysteretic effects.  \cite{Hristov2018,Hristov2019a}
	Throughout this paper we focus only on the real part of the elastoresistivity signal, and neglect the small, but finite, imaginary component.
	Understanding the physical effects producing the quadrature signal would require a thorough investigation of the frequency dependence, which has not been conducted as part of this work.

\section{Results}\label{sec:results}

	We now describe the various experimental results on \rte{} under uniaxial stress, each of which provides further support of a strain-induced CDW realignment.
	We begin in \cref{sec:ece} by probing the upper phase transition with the elastocaloric effect.
	Here, the reversal of a thermodynamic anomaly suggests a reversed anisotropy of the thermal expansion tensor, which we link to the CDW order parameter itself.
 	\Cref{sec:resistivity} then describes the effect of uniaxial stress on all three components of the resistivity tensor.
 	Here we show that while $\rho_{bb}$ suggests the presence of a CDW gap regardless of stress, stress strongly suppresses gap-related effects in $\rho_{aa}$.
 	Finally, \cref{sec:er} examines the elastoresistivity, and confirms that the antisymmetric, in-plane component of strain controls the strain-induced transition.

	\subsection{Thermodynamic evidence of CDW switching: elastocaloric effect}\label{sec:ece}

		\begin{figure*}
			\centering
			\includegraphics[width=\textwidth]{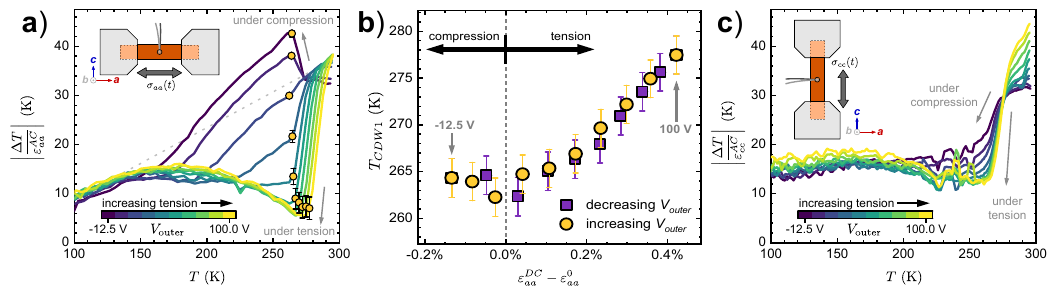}
			\caption{
				Elastocaloric effect results for a sample of \er{} with uniaxial stress
				(a) Magnitude of the elastocaloric response with stress parallel to the \aax{} as a function of temperature and offset voltage.
				The critical temperature \ti{} (circles) is extracted from an extremum in the second derivative.
				Inset: schematic of the experimental setup.
				The dotted line is a guide to the eye for a linear background independent of strain.
				The anomaly at \ti{} rises above this background for compressive stresses, but falls below this background for large enough tensile strain.
				(b) Critical temperatures \ti{} extracted from (a) as a function of DC offset strain, shown for both increasing and decreasing PZT voltage steps.
				The critical temperature rises as one departs from $\varepsilon_{aa}^0$ toward either compressive or tensile strain.
				(c) Elastocaloric effect with stress parallel to the \cax{}.
				Tension increases the height of the critical anomaly, but no switching behavior is observed.
			}
			\label{fig:erte3_ece}
		\end{figure*}

%		\paragraph{ECE description}
		The elastocaloric effect (ECE) reflects the strain-dependence of the entropy of a material.
		This can be detected experimentally through the change in temperature resulting from an adiabatic change in strain: $(dT/d\varepsilon)_S$.
		Near phase transitions, ECE measurements manifest distinct anomalies, similar to those seen in heat capacity, due to strain-dependence of the critical temperature.\cite{Hristov2019}
		More generally, though, the ECE is related to specific components of the elastic stiffness tensor, the thermal expansion tensor, and the heat capacity.
		Before discussing these relations in greater detail, we first present the phenomenology uncovered by the ECE measurements of \rte{}.

\begin{figure}
			\centering
			\includegraphics[width=0.8\columnwidth]{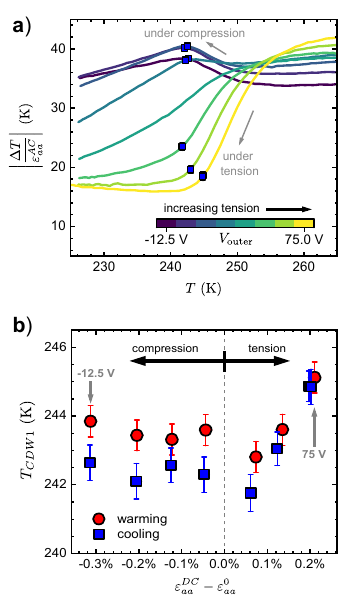}
			\caption{
				Elastocaloric effect measurements in \tm{} with stress parallel to the \aax{}.
				(a) Magnitude of the elastocaloric response near \ti{} as a function of temperature and PZT voltage.
				Each trace was collected by cooling at a constant $V_{outer}$, and \ti{} is extracted by the second derivative.
				Despite using different data acquisition protocols and different rare earths, the results in \cref{fig:erte3_ece} share the same behavior.
				(b) Extracted critical temperatures on the same sample for both warming and cooling traces.
				The increase of \ti{} for tensile strains is clearly defined, although an increase on the compressive side is not observed.
				We attribute this to slight buckling of the sample, resulting in an overestimate of the compressive strain.
				One of the curves near the center exhibited no clear extremum in the second derivative, so \ti{} could not be unequivocally identified.
				Tension and compression are labeled (black arrows) relative to $\tilde{\varepsilon}_A=0$.
			}
			\label{fig:tmte3_ece}
		\end{figure}

		The elastocaloric effect (ECE) in \er{} is presented in \cref{fig:erte3_ece}, and similar data for \tm{} are shown in \cref{fig:tmte3_ece}.
		Panel (a) in both
		\cref{fig:erte3_ece,fig:tmte3_ece} corresponds to stress parallel to the \aax{}.
		For slightly compressive stresses (negative $V_\text{outer}$), a steplike anomaly at \ti{} causes the ECE to increase in magnitude upon cooling through the transition (gray arrow marked ``under compression'').
		Under these conditions, \ti{} corresponds to the onset of the \qc{} phase as in freestanding crystals.
		Increasing tensile strain causes the step first to shrink, then to invert such that the ECE decreases in magnitude upon cooling through \ti{} (gray arrow marked ``under tension'').
		At the largest tensile strains, the transition into the CDW state corresponds to an $\approx$85\% decrease of the total ECE signal.

%		\paragraph{ECE definitions}
	 	We plot the strain dependence of \ti in \cref{fig:erte3_ece}(b).
	 	We define $\varepsilon_{aa}^0$ as the strain for which \ti{} reaches a minimum, corresponding to $\tilde{\varepsilon}_A=0$ in Section \ref{sec:landau}.
	 	This point, where $dT_{CDW1}/d\varepsilon$ changes sign, also corresponds to where the step-like anomaly flips its direction.

%		\paragraph{ECE:C-axis description}
		Such switching behavior is not observed for tensile stress parallel to the \cax{}, as seen for \er{} in \cref{fig:erte3_ece}(c).
		In contrast to \aax{} stress, even the most compressive traces correspond to a decrease in ECE magnitude upon cooling through \ti{}.
		Additionally, increasing tensile strain slightly increases \ti{}.
		For $T\lesssim150~\text{K}$ in both orientations, the ECE curves again become largely independent of both the orientation and magnitude of the stress.

%		\paragraph{TmTe3 ECE}
		ECE measurements in \tm{} for \aax{} strain, shown in \cref{fig:tmte3_ece}, follow the same trends as \er{}.
		Cooling through \ti{} ($\approx245$~K) increases the ECE under compressive stress, and decreases the ECE under tensile stress.
		\ti{} also exhibits a weak minimum for small tensile stress.
		We attribute the strain-independence of \ti{} on the compressive side to buckling of the sample.
		Buckling would imply poor transmission of strain into the sample, consistent with the smaller ECE signal observed for the most compressive case, ${V_\mathrm{outer}=-12.5}$~V.

		Each curve in \cref{fig:tmte3_ece} was taken on a separate temperature sweep with a constant voltage offset $V_\mathrm{dc}$.
		The data in \cref{fig:erte3_ece}, in contrast, was all taken on the same temperature sweep, but stepping the voltage as in \cref{fig:experiment}(d).
		Both, however, display the same qualitative features, indicating that this behavior is a robust feature of \rte{} and does not depend on the stress and temperature history.

		%% DISCUSSION %%
		The change in behavior in the ECE between \aax{} tension and compression can in principle arise from two separate physical effects.
		Firstly, at temperatures near a continuous phase transition, the strength of the critical fluctuations depends on the reduced temperature $t=(T-T_c)/T_c$, where $T_c$ is the critical temperature.
		If $T_c$ is tuned adiabatically by an external parameter such as strain, the sample temperature will shift such that the total entropy is conserved.
		The smaller $|t|$, the larger the change in temperature.
		The resulting elastocaloric effect is proportional to the critical part of the specific heat $C_p^{(c)}$ and the rate of change of $T_c$ with strain \cite{Ikeda2019,Hristov2019}
		\begin{equation}\label{eq:fisherlanger}
			\left(\frac{dT}{d\varepsilon_{ij}}\right)_S = \frac{C_p^{(c)}}{C_p} \frac{dT_c}{d\varepsilon_{ij}} + \cdots
		\end{equation}
		In \rte{}, we see that the change in sign of the ECE step occurs at the same strain as the change in sign of $dT_{CDW1}/d\varepsilon$, consistent with this relation.
		
		Heat capacity anomalies at \ti{} have been observed to be quite small ($\approx 1 \%$ of the total $C_p$) at both of the CDW transitions in TbTe\tri{}\cite{Saint-Paul2016} and \er{}\cite{Saint-Paul2017}.
		The derivative of \ti{} with respect to $\varepsilon_{aa}$ is approximately 20~K/\%.
		Using this value, \cref{eq:fisherlanger} predicts an ECE anomaly at \ti{} of approximately 20~K (0.2~K/\%) which is indeed what is observed in both \cref{fig:erte3_ece,fig:tmte3_ece}.
		In \rte{}, neither of the CDW transitions can be considered mean-field\cite{Yao2006,Banerjee2013,Saint-Paul2016}, so some deviation from a mean-field step is expected.

		%	\paragraph{ECE+Thermal Expansion}
		This effect, however, only applies in the fluctuation regime within 10-20~K near \ti{}.\cite{Eiter2013}
		\Cref{fig:erte3_ece} shows that the sign-changing behavior of the ECE anomaly (subtracting a linear background) spans almost 100~K.
		Well below the transition, the strain-dependent ECE must therefore arise from the equilibrium CDW phase itself, rather than from critical fluctuations.
		A more general expression for the elastocaloric effect, valid at any temperature, relates the ECE to several thermodynamic properties
		\begin{equation}
			\left(\frac{dT}{d\varepsilon_{ij}}\right)_S= \frac{-T}{C_\sigma}  C_{ijkl} \alpha_{kl}
		\end{equation}
		where $C_{\sigma}$ is the specific heat at constant stress, $C_{ijkl}$ is the elastic constant tensor, and $\alpha_{kl}$ is the thermal expansion tensor.
		Regardless of the strain in the sample, stability requires that $C_\sigma$ and $C_{ijkl}$ must retain the same sign, but $\alpha_{kl}$ has no such constraint.
		Therefore, a strain-induced change in the thermal expansion coefficients would explain the behavior of the ECE in \rte{}.

		Previous x-ray measurements\cite{Ru2008b, Sacchetti2009} indicate that the antisymmetric in-plane thermal expansion $\alpha_A = (\alpha_{cc} - \alpha_{aa})/2$ is negative below \ti{}.
		Ordering of the \qc{} CDW therefore reinforces the built-in orthorhombic distortion, increasing the $c$ lattice parameter relative to $a$.
		The observed change in sign in the elastocaloric anomaly, and therefore in the thermal expansion, suggests that the CDW wavevector has been switched from parallel to the \cax{} to parallel to the \aax{}.

	\subsection{Tracking changes in Fermi surface anisotropy: resistivity}\label{sec:resistivity}

		\begin{figure}
			\centering
			\includegraphics[width=\columnwidth]{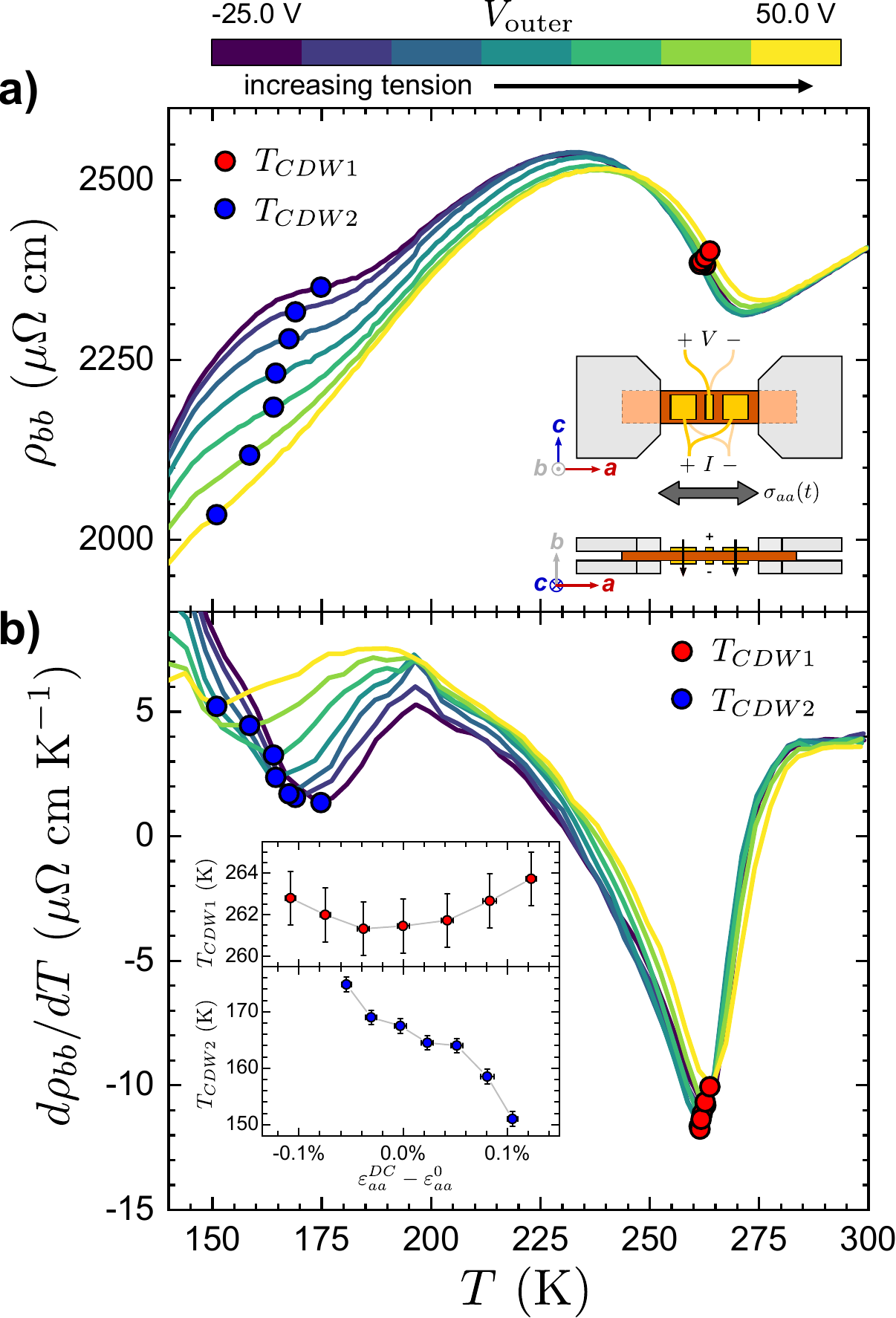}
			\caption{
				Out-of-plane resistivity $\rho_{bb}$ in \er{}.
				Panel (a) shows the resistivity traces for a series of PZT voltages.
				Inset: top and side view schematics of the contact geometry used in this experiment.
				Current is passed into the sample through the larger, outer two contacts on both the top and bottom of the crystal, and the smaller center pads are used to detect the voltage.
				Temperature derivatives of the resistivity traces are shown in panel (b).
				The minima of the derivative traces correspond to the critical temperatures.
				Inset to (b): \ti{} and \tii{} plotted as a function of strain.
				The lack of a maximum in \tii{} corresponding to the minimum in \ti{} and the switch between the two cases is attributed to an underestimate of the tensile stress arising from mismatched thermal expansion of the sample and stress cell.
			}
			\label{fig:rhobb}
		\end{figure}

%		\paragraph{Out of Plane}
		In this section, we describe measurements of all three components of the orthorhombic resistivity tensor as a function of temperature and applied stress.
		Electrical transport is highly sensitive to the opening of CDW gaps.
		These measurements show similar switching behavior as the ECE.
		Specifically, we show that a gap still opens in the strain-induced state, but at a different location on the Fermi surface, as would be expected from a CDW reorientation transition.

		Previous results\cite{Sinchenko2014,Walmsley2017} have established that the onset of the first and second CDWs, with wavevectors directed along the \textit{c-} and \aax{} respectively, invoke a larger change in $\rho_{aa}$ and $\rho_{cc}$ respectively.
		This seemingly counterintuitive result arises due to the variation of the Fermi velocity around the Fermi surface (FS), in conjunction with which sections of the FS are gapped at each CDW transition.
		In contrast, the longitudinal resistivity for currents running perpendicular to the Te planes, $\rho_{bb}$, is equally sensitive for both CDW transitions.
		Measurements of $\rho_{bb}$ for \er{} under \aax{} stress are presented in \cref{fig:rhobb}.
		\ti{} and \tii{} are identified as minima in the temperature derivative.

		We find that \ti{} shows the same trend as ECE measurements--a weak minimum for small tensile stress.
		Combining the trend in \ti{} with the ECE and resistivity measurements, we can conclude that the sample undergoes a phase transition which reorients the CDW wavevector and gap.
		Specifically, the height of the resistivity increase below \ti{}, which is related to changes in the density of states at the Fermi level due to the opening of the CDW gap, decreases only slightly as stress increases.
		Due to the complex way in which the size, shape, and location of the CDW gap affects the electrical resistivity, the magnitude of the bump below \ti{} should not be taken as a quantitative measure of the CDW gap.
		However, the persistence of a sharp peak in the derivative of $\rho_{bb}$ suggests that a gap always opens somewhere on the quasi-2D FS.
		This suggests that the phase transition observed in ECE and in the trend of \ti{} is a reorientation of the CDW wavevector.
		
		\tii{} decreases significantly as tensile strain increases, dropping $\approx$20~K for strains of 0.1\%.
		Simultaneously, the corresponding bump in resistivity decreases in magnitude as well.
		Unlike \ti{}, \tii{} decreases monotonically with increasing tensile strain.
		We attribute this to an underestimate of the true tensile strain arising from mismatched thermal expansion between sample and the stress cell.
		The monotonic decrease in \tii{} suggests that the tensile stress is, at these temperatures, beyond that required to realign the primary wavevector by 90\degree.
		The effects of thermal expansion mismatch is discussed in detail in \cref{app:nonideal}.
		This effect is also clearly visible in \cref{fig:tmera}(b).

		\begin{figure}
			\centering
			\includegraphics[width=\columnwidth]{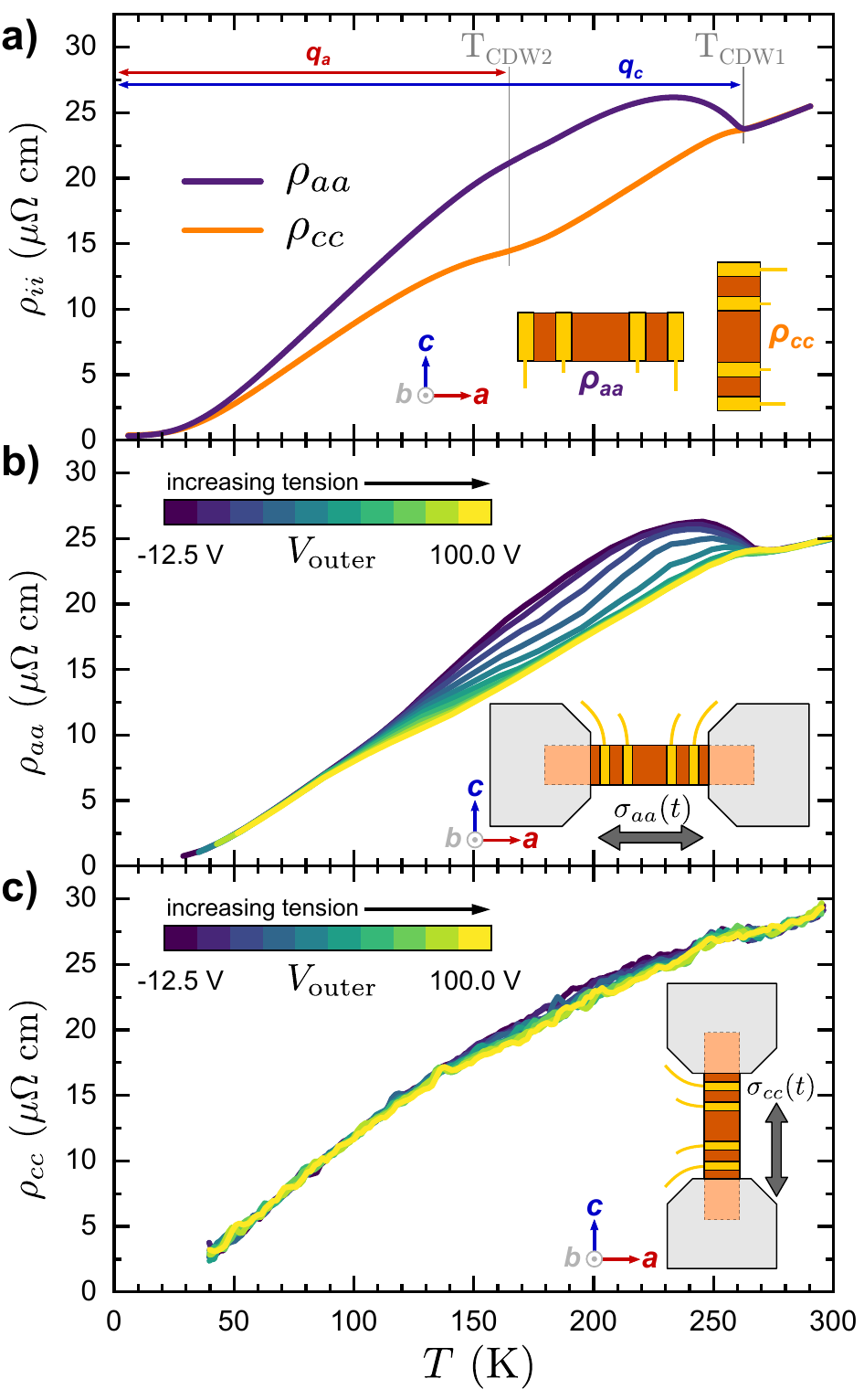}
			\caption{
				In-plane resistivity components of \er{} as a function of temperature and strain.
				(a) free-standing resistivity components of \er{}.
				Resistivity along the \aax{} increases more strongly upon entering the \qc{} state due to the curvature of the gapped region of the Fermi surface\cite{Sinchenko2014}.
				(b) \aax{} resistivity of \er{} under \aax{} stress, as a function of temperature and PZT voltage.
				Negative and low voltages behave similarly to the freestanding case for $\rho_{aa}$, but crosses over to resemble freestanding $\rho_{cc}$ as the tension increases.
				(c) \cax{} resistivity under \cax{} stress, which does not demonstrate any switching behavior.
			}
			\label{fig:rhoinplane}
		\end{figure}

%		\paragraph{IN PLANE}
		The in-plane resistivity components, $\rho_{aa}$ and $\rho_{cc}$, are presented in \cref{fig:rhoinplane} for \er{}.
		These components show similar features at the CDW transitions, but also provide insight into the anisotropy of the Fermi surface in the ordered phases.
		As mentioned before, the CDW gap with wavevector \qc{} has a larger effect on $\rho_{aa}$ than on $\rho_{cc}$, and vice versa.
		This is observed clearly in the top panel of \cref{fig:rhoinplane} for a freestanding sample.

		The lower two panels in \cref{fig:rhoinplane} show measurements of both $\rho_{aa}(\varepsilon)$ and $\rho_{cc}(\varepsilon)$ with uniaxial stress parallel to the current flow.
		For slightly compressive \aax{} stress, $\rho_{aa}(\varepsilon)$ closely resembles the freestanding value, $\rho_{aa}(\varepsilon=0)$.
		As $\sigma_{aa}$ increases, however, the change of $\rho_{aa}$ across \ti{} decreases in magnitude, and $\rho_{aa}(\varepsilon_{aa})$ eventually behaves like $\rho_{cc}(\varepsilon=0)$ instead.
		In contrast, no such switching behavior is seen in $\rho_{cc}(\varepsilon)$.
		A slight decrease of the resistivity can be observed for the largest tensile stresses, but the curves never deviate very far from $\rho_{cc}(\varepsilon=0)$.

		Tensile $\sigma_{cc}$ reinforces the intrinsic orthorhombicity of the material.
		Tensile $\sigma_{aa}$, however, opposes this orthorhombicity.
		In the simplified model presented in Section \ref{sec:landau}, tensile $\sigma_{aa}$ can train the CDW wavevector along \qa{}.
		Our observation of clear switching behavior in $\rho_{aa}$, coupled with the persistence of a CDW gap in $\rho_{bb}$ and a minimum in \ti{}, support this interpretation, and suggest a strain-tuned reorientation transition between \qa{} and \qc{} states.
		It is likely that compressive $\sigma_{cc}$ would also cause a reorientation, although sample buckling makes this regime inaccessible in the present experiment.

	\subsection{Evidence for first-order strain-induced transition: hysteresis in resistivity}\label{sec:hyst}

		\begin{figure}
			\centering
			\includegraphics[width=\columnwidth]{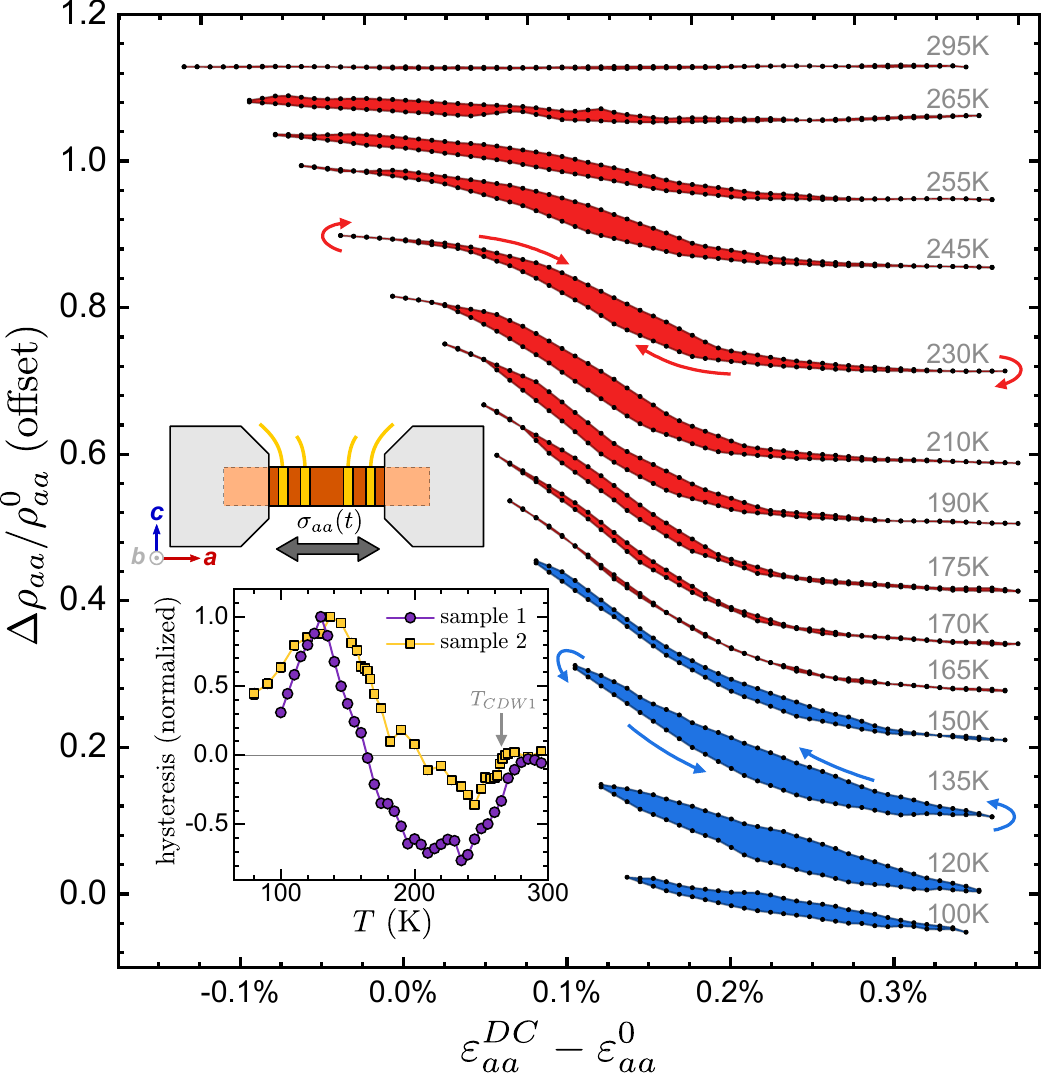}
			\caption{
				Ratiometric change in $\rho_{aa}$ of \er{} through stress cycles performed at a few representative temperatures.
				Curves are offset vertically for clarity.
				Inset: Schematic of the contact and stress geometry, and plot of the integrated area contained within the hysteresis loops in (b) for two different samples of \er{}.
				In both, the hysteresis loops open as the sample cools through \ti{}.
				The hysteresis loop changes direction (clockwise to counterclockwise) as the temperature decreases further.
				The extracted area is sensitive to the exact DC strain range accessed, which accounts for the differences between samples.
				The observed hysteresis clearly indicates the first-order nature of the phase transition where the direction of the CDW wavevector switches from \qa{} to \qc{}, corresponding to the schematic phase diagram shown in \cref{fig:glrte3}(f).
			}
			\label{fig:hyst}
		\end{figure}

		In order to examine the character of this putative reorientation transition, we have performed $\sigma_{aa}$ stress cycles on a sample of \er{} at constant temperature while measuring the \aax{} resistivity, presented in \cref{fig:hyst}.
		Above \ti{}, very little strain-induced change in the resistivity is observed.
		As the sample cools below \ti{}, a clear hysteresis loop opens.
		The exact size and shape of the hysteresis loops depends on the extent of the stresses applied, which depend on the history of the PZT actuators and is not identical between samples.
		The inset of \cref{fig:hyst} shows the normalized loop area for two different \er{} samples in the same configuration.
		This hysteretic behavior below \ti{} demonstrates that the switching behavior seen in both the elastocaloric and resistivity data corresponds to a first order transition.

		Further decreases in temperature cause the hysteresis to close again, then reopen with the path oriented in the counterclockwise direction instead of clockwise.
		This inversion occurs near \tii{}, although the exact value varies between the two samples.
		(Stress-cycle data for the second sample is shown in \cref{sec:hyst2}.)
		Practical limitations in the stress cell (as described in \cref{app:nonideal}) caused the two measurements to cover slightly different ranges of strains--this in turn affects the details of the hysteresis loop shape and size.
		However, the presence of hysteresis, as well as the sign-changing behavior, appear to be robust.

		The existence of a first-order transition and the vanishing hysteresis at a temperature below \ti{} are both consistent with the sixth-order free energy expansion of Section \ref{sec:landau} and \cref{fig:glrte3}(f).
		\Cref{eq:fe2} produces two multicritical points at the intersections between the first-order transition and \ti{} and \tii{}.
		At either point, the transition between a \qc{} and \qa{} state must become continuous because the tunneling barrier will vanish.

		The model of \cref{eq:fe2} does not, however, explain our observation of hysteresis for $T\lesssim T_{CDW,2}$.
		At low temperatures, the model exhibits a single bidirectional CDW phase, free of any first order transitions.
		Our measurements in \Cref{fig:hyst} suggest, however, that a first-order transition does exist below \tii{}.
		This low temperature hysteresis may arise from subtle differences between the \qa{} and \qc{} phases; for instance, the magnitudes $|\mathbf{q_c}|$ and $|\mathbf{q_a}|$ differ slightly in freestanding samples\cite{Banerjee2013,Kogar2020}.
		This effect is neglected in \cref{eq:fe2}, but details like this could introduce additional first-order transitions to the phase diagram.
		In any case, further experiments are required to fully explain these observations.

	\subsection{Disentangling in-plane symmetric and antisymmetric strains: Elastoresistivity}\label{sec:er}
		\begin{figure*}
			\centering
			\includegraphics[width=\textwidth]{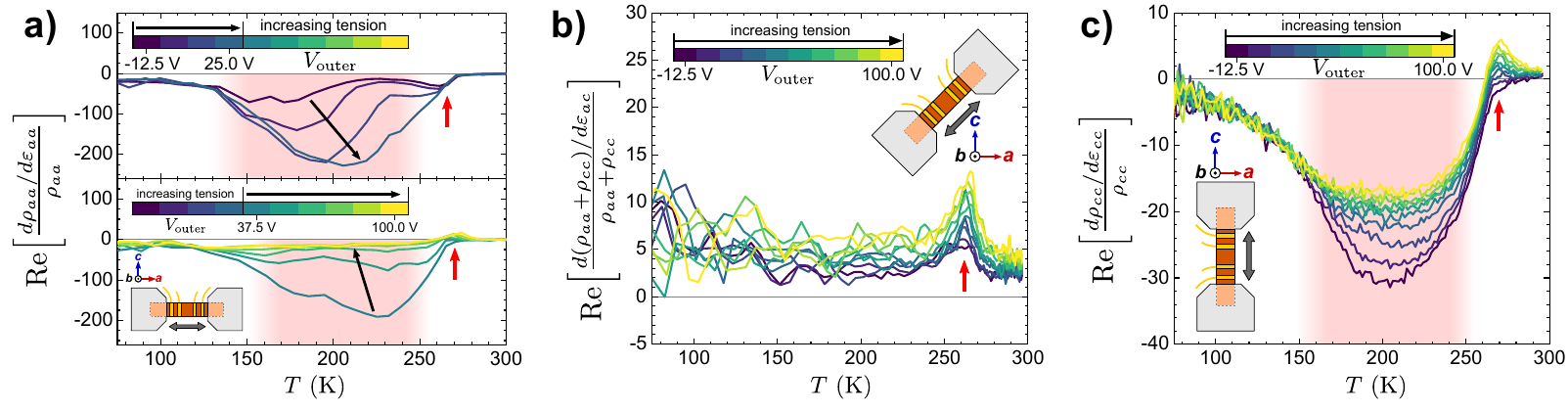}
			\caption{
				(a)-(c) Longitudinal elastoresistivity components for \er{} under three different stress orientations, as functions of temperature and PZT voltage.
				The results from the \aax{} sample are split into two subpanels for clarity.
				Insets describe the orientation of the sample and current in each experiment.
				In both the \aax{} case, increasing tensile strain initially causes the growth of a large negative bulge (shaded in pink) between approx. 150~K and 265~K, which then increases again for PZT voltages 37.5~V and above.
				The behavior in the bottom panel of (a) mimics that of (c), where tensile \cax{} stress also decreases the magnitude of this negative bulge in the same temperature range.
				Diagonal stress, as shown in (b), does not generate any such response, indicating that the antisymmetric strain component $\varepsilon_{aa}-\varepsilon_{cc}$ dominates the changes in resistivity.
				A peak or dip localized at \ti{} (shown by the red arrow) appears in all three sample orientations although it carries the opposite sign for compressive and small tensile \aax stress in the top subpanel of (a).
				This must therefore arise from the coupling of the CDW order parameter fluctuations to the in-plane symmetric strain or out-of-plane strain.\cite{Hristov2019}
			}
			\label{fig:erac}
		\end{figure*}

		Finally, we turn to an elastoresistivity technique\cite{Hristov2018} to distinguish the effects of different strain components on the CDW state.
		The elastoresistivity (ER) tensor $m_{ijkl}$, which relates normalized changes in the resistivity tensor $\rho_{ij}$ to the material strain $\varepsilon_{kl}$ is defined as
		\begin{equation}
			m_{ijkl} = \frac{d(\Delta\rho/\rho_0)_{ij}}{d\varepsilon_{kl}}
		\end{equation}
		where $\delta\rho$ is the strain-induced change in resistivity and $\rho_0$ is the resistivity under zero strain conditions.  		\footnote{The normalizing resistivity $\rho_0$ is usually taken as the geometric mean of the relevant resistivity tensor components at zero strain.\cite{Shapiro2015}
			However, due to large ($\approx20\%$) changes in the resistivity as shown in \cref{fig:rhoinplane}(b), we elect instead to normalize by the simultaneously measured average resistivity.
			This allows us to present our data without making any ancillary measurements or extrapolations.}
		ER has proven a powerful tool in the understanding of symmetry-breaking phase transitions.\cite{Chu2012,Riggs2014,Kuo2016,Rosenberg2019}
		\Cref{fig:erac} shows the in-plane, longitudinal ER responses of \er{} under three different in-plane stress conditions.

		Stress and current aligned with the \aax{} generates the largest ER as well as the largest stress dependence.
		We identify two persistent features in the data--the first, near \ti{}, is a sharp dip toward negative ER values (red arrow in \cref{fig:erac}(a)).
		Increasing tensile stress causes this feature to change sign, becoming a peak.

		The second feature is a broad region (shaded in pink) of large negative ER.
		Increasing the tensile strain causes this ``bulge'' to increase in magnitude, then recede again.
		The largest magnitude (ER$\approx-225$) is reached for the same stress at which the first feature changes sign.

		In contrast, \cax{} longitudinal ER (\cref{fig:erac}(c)) only shows behavior similar to the large stress limit of the \aax{} ER.
		The same two features are present: a small peak near \ti{} (red arrow) and a broad minimum (shaded region) which shrinks with increasing tension.

		Finally, the longitudinal ER response aligned 45\degree{} to the orthorhombic axes (\cref{fig:erac}(b)) shows very different behavior.  The small positive peak near \ti{} is still present and increases in magnitude with tension.
		Below \ti{}, however, the ER response is flat, small, and relatively insensitive to stress.

		\begin{figure}
			\centering
			\includegraphics[width=\columnwidth]{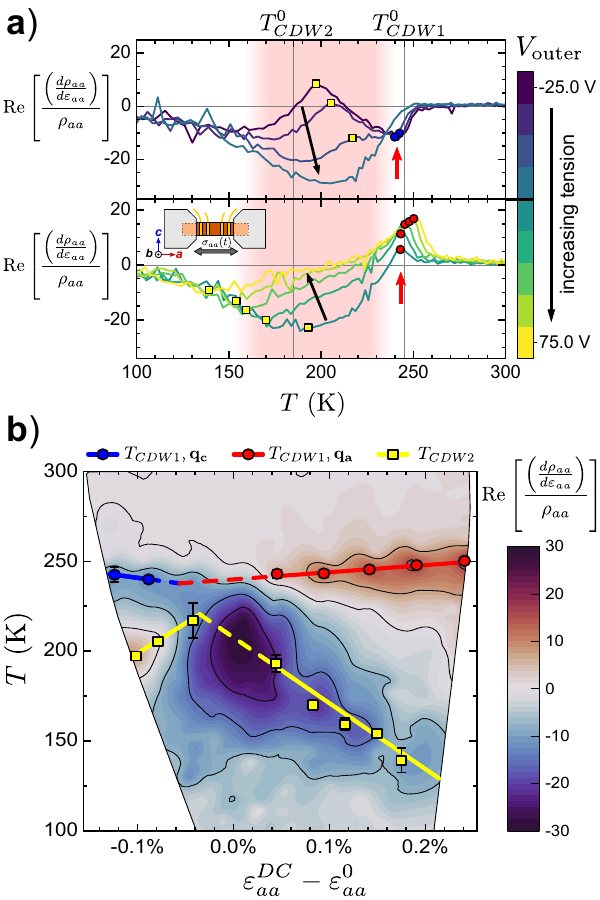}
			\caption{
				Longitudinal \aax{} elastoresistivity in \tm{}.
				Data presented here was taken by stepping the PZT voltage while sweeping temperature and extracted for steps of increasing tension only.
				(a) Elastoresistivity as a function of temperature and PZT voltage.
				Traces are all from the same sample, but have been separated into two panels for clarity.
				The top panel contains the responses for the most compressive/least tensile stress conditions, and the bottom for the largest tensile stresses.
				Short red arrows point out the peaks and dips attributed to \ti{}, and the shaded regions highlight the ``bulge'' feature.
				(b) The same data as in (a) presented in the strain-temperature plane.
				Symbols correspond to the extracted extrema in the traces in (a).
				Solid lines are linear fits to the extracted transition positions and dotted lines denote an extrapolation to the intersection points.
				In conjunction with the evidence for a first order transition as shown in \cref{fig:hyst}, this diagram maps most closely onto the phase diagram presented in \cref{fig:glrte3}(f).
			}
			\label{fig:tmera}
		\end{figure}

		The orthorhombic symmetry group \textit{Cmcm} imposes very few constraints on the coupling between the CDWs and uniform strain $\varepsilon_{ij}$
		Consider a generalized coupling of the form $\varepsilon_{ij}\Lambda_{ijk}|\psi_k|^2$, where $k=a,c$.
		Mirror planes impose $\Lambda_{ijk}=0$ for $i\ne j$, but the 6 other terms remain free.
		In the tetragonal limit, one may take $\Lambda_{iik}=\Lambda_{kki}$.
		However, as uniaxial stress experiments simultaneously tune several $\varepsilon_{ii}$ components, disentangling the effects of each on the CDW is not trivial.

		However, the three ER measurements presented in \cref{fig:erac} allow us to identify the in-plane antisymmetric component of the strain, $\varepsilon_A=(\varepsilon_{aa}-\varepsilon_{cc})/2$, as the operative tuning parameter for the strain-induced transitions.
		Uniaxial stresses which do not generate a nonzero $\varepsilon_A$ do not induce a transition.
		Furthermore, the in-plane symmetric component $\varepsilon_S=(\varepsilon_{aa}+\varepsilon_{cc})/2$ and the out-of-plane component $\varepsilon_{bb}$ only contribute to shifts in critical temperature.

		To see this, consider first the $a$- and $c$-axis elastoresistivity measurements.
		Both show similar phenomenology below \ti{} under large tensile stress.
		Except for a small anisotropy in the elastic constants, these measurements are performed under similar magnitudes of $\varepsilon_S$, $\varepsilon_{bb}$ and $\varepsilon_A$, although the sign of $\varepsilon_A$ will be flipped.
		Applying uniaxial stress along the in-plane diagonal results in $\varepsilon_A\approx 0$ and $\varepsilon_{ac}\ne0$, although shear strain cannot couple (to linear order) to the CDW.
		The symmetric components $\varepsilon_S$ and $\varepsilon_{bb}$ will again be similar.
		This diagonal configuration effectively only probes the response to symmetric strains.

		The large negative ER response present below \ti{} in the $a$- and $c$-axis configurations can therefore only arise from coupling to $\varepsilon_A$.
		Meanwhile, the feature near \ti{}, observed in all three measurements, must arise from coupling to symmetric strains $\varepsilon_S$ and $\varepsilon_{bb}$.
		Near a strain-tuned continuous phase transition, ER will display similar singular behavior to the heat capacity.\cite{Hristov2019}
		We can therefore attribute this feature to changes in \ti{} caused by $\varepsilon_S$ and $\varepsilon_{bb}$.

		Similar phenomenology is also seen for the \aax{} ER in \tm{} (\cref{fig:tmera}(a)).
		A sharp feature (red arrow) is observed near \ti{}, as well as a broad minimum in ER at lower temperatures.
		A critical value of stress ($V_\text{outer}\approx +12.5$) corresponds to both a change in sign of the peak at \ti{}, and the largest negative ER.

		Additionally, a sharp local maximum is observed at low temperatures on the compressive side.
		On the tensile side, a similarly sharp minimum is observed--both of these features, given their opposing strain dependence, can be identified with \tii{} and a transition into a bidirectional CDW state.
		With the exception of the most compressed stress values, \tii{} shows a similar monotonic decrease with stress\cref{fig:rhobb}(b).

		\Cref{fig:tmera}(b) compiles the strain- and temperature-dependent evolution of the ER in \tm{}, as well as the extracted transition temperatures.
		Lines of best fit to \ti{} and \tii{}--the intersections of which highlight the likely positions of two separate multicritical points--highlight the fact that the phase diagram displays the same topology as that produced in Fig.~1(f) by the free energy expansion.
		Combined with the evidence for a first order transition, the resulting phase diagram appears consistent with \cref{fig:glrte3}(f).

		%\paragraph{Transition slopes}
		\begin{table}
			\caption{
				Rate of change of the critical temperatures \ti{} and \tii{} with strain in the strain-induced \qa{} state.
				In all of these experiments, uniaxial stress is applied long the \aax{}.
				Not all techniques are sensitive to \tii{}.
				The strain range of the $\rho_{bb}$ measurements was insufficient to observe consistent linear behavior in \ti{}.
				Sources of sample-to-sample variation are described in the text.
			}\label{tab:dtcde}
			\begin{tabularx}{\columnwidth}{cC{1.3cm}YY}
				\toprule
				& & $\displaystyle\frac{dT_{CDW1}}{d\varepsilon_{aa}}$~(K/\%) & $\displaystyle\frac{dT_{CDW2}}{d\varepsilon_{aa}}$~(K/\%) \\ [1.5ex]\midrule
				\multirow{3}{*}{\er{}}
				& ECE          & $47 \pm 11$             & \multicolumn{1}{c}{$\cdot$} \\
				& $\rho_{aa}$  & $68.4 \pm 6.7$              & \multicolumn{1}{c}{$\cdot$} \\
				& $\rho_{bb}$  & \multicolumn{1}{c}{$\cdot$} 				& $-130 \pm 16$          \\
				\midrule
				\multirow{3}{*}{\tm{}}
				&	ECE          & $22.1 \pm 5.8$  &  \multicolumn{1}{c}{$\cdot$}  \\
				& $\rho_{aa}$  & $51.5 \pm 6.7$  &  \multicolumn{1}{c}{$\cdot$}  \\
				& ER           & $43.7 \pm 9.8$  &  $-221 \pm 39$            \\
				\bottomrule
			\end{tabularx}
		\end{table}

		The change in critical temperatures for the strain-induced \qa{} state under strain are presented in \cref{tab:dtcde}.
		In general, \tii{} is seen to be approximately 3-5 times more sensitive to strain than \ti{}.
		The relatively large scatter in the different measurement techniques likely arises from uncertainty in the strain transmission from the stress cell to the sample, as well as run-to-run variations tentatively attributed to differences the thickness of the epoxy layer used for mounting and uncertainties in the spacing between the mounting plates.
		Additionally, near the critical point the predicted sharp cusp in \ti{} is seen to be rounded somewhat, likely due to strain inhomogeneities in the sample.
		This curvature can also lead to an underestimate of $dT_{CDW1}/d\varepsilon_{aa}$.
		However, using a Young's modulus $E$ of approximately 50~GPa\cite{Saint-Paul2019}, these values agree with the range of values of $dT_{CDW1}/d\sigma_{aa}=E^{-1}(dT_{CDW1}/d\varepsilon_{aa})$ previously reported for several \rte{} compounds.\cite{Saint-Paul2016,Saint-Paul2017,Saint-Paul2018}

\section{Discussion}\label{sec:disco}

% Twinning
	{A phase transition induced by in-plane, antisymmetric strain in a weakly-orthorhombic material raises a natural question of the existence of twin domains, either naturally occurring or induced by the applied stress.
	In the case of \rte{}, the most prevalent naturally occurring defect is a planar stacking fault normal to the $b$-axis, at which the $a$- and $c$-axes swap directions.
	The existence of one or more of such stacking faults would tend to average the $a$- and $c$-axis observables together; we take the strikingly different behavior for stress along the $a$ and $c$ axes (\cref{fig:erte3_ece}(a) and (c), or \cref{fig:rhoinplane}(b) and (c), for example) as evidence that such stacking faults and misaligned domains are either not present or are sufficiently rare as to have negligible effect.
	Furthermore, macroscopic strains can neither couple to nor switch the direction of the glide plane itself due to the non-symmorphic nature of the crystal structure.
	In order to swap the symmetries of the $a$ and $c$ planes, one would need to shift every other $R$Te block layer by $(a/2, 0, c/2)$.
	In-plane antisymmetric strain can only alter the dimensions of the unit cell, and the existence of any microscopic orthorhombic domains must rely only on the direction of the CDW ordering vector, not the crystal structure.}

%	Future directions in this material alone

	This work immediately suggests the application of other standard techniques to \rte{} under uniaxial stress.
	X-ray diffraction, angle-resolved photoemission spectroscopy (ARPES), and pump-probe reflectivity or diffraction measurements under stress would be of particular interest.
	Diffraction measurements would enable the direct measurement of the intensity, position, and correlation length of the superlattice peaks at \qa{} and \qc{} simultaneously with the lattice strain.
	ARPES would provide direct verification and quantification of the CDW gap at \qa{}.
	Other information could be gained through various ultrafast pump-probe techniques under strain as well; recent work has demonstrated that laser pulses on freestanding LaTe\tri{} can induce both topological defects in the CDW phase\cite{Zong2019,Zong2019a} and transient \qa{} CDW correlations.
	These tools would provide a deeper understanding of the dynamics of the strain-induced \qa{} state.

% Relating these results to cuprates

	On a broader scale, \rte{} offers a platform for systematic studies of the effects of chemical disorder on CDW-strain interactions.
	Several recent works\cite{He2016,Lou2016,Straquadine2019b} have shown that Pd atoms inserted between the Te bilayers (\pdx) act as a weak disorder potential which frustrates and suppresses long-range CDW order.
	Examination of the CDW correlations in \pdx{} with scanning tunneling microscopy (STM)\cite{Fang2019} suggests that increasing disorder may cause a cascade of ``melting'' transitions; the long-range-ordered CDW gives way first to a Bragg glass \cite{Giamarchi1995}, then a vestigial nematic phase\cite{Nie2014,Nie2017}, before finally losing all long-range coherence for sufficiently strong disorder.
	Both of these intermediate phases break rotational symmetry and should therefore couple to in-plane antisymmetric strain.
	Uniaxial stress experiments such as those presented in this work, but applied to \pdx{}, would enable both the substantiation of this phase diagram as well as an in-depth study of the properties of these phases and associate phase transitions.
	\rte{} and \pdx{} may be one of the few material systems in which the physics of these states can be examined systematically.

	Expanding our focus beyond the \rte{} family itself, understanding gained in experiments of strain- and disorder tuned \rte{} would also be particularly relevant for the case of the high-$T_c$ cuprates, where disorder is an unavoidable side effect of chemical doping.
	Short- and long-range charge-ordered phases only appear in the cuprates at finite doping levels, but \rte{}, as a model system, may provide experimental access to an otherwise inaccessible limit of charge order on a square lattice in the absence of disorder.
	Having a chemically ``clean'' model system such as \rte{} allows the experimenter to continuously track the evolution of symmetry breaking effects (such as CDW correlations) and their sensitivity to in-plane antisymmetric strain as disorder is increased.
	In this way it may be possible to identify observed effects in disordered systems as the smeared, frustrated vestiges of phase transitions in the clean limit.

\section{Conclusions}\label{sec:conc}
	In this work, we have shown that \rte{}, specifically \er{} and \tm{}, provide a practical model system for the study of strain-CDW interactions.
	In particular, uniaxial stress can tune \rte{} to and beyond a quasi-tetragonal state for which the CDW wavevector realigns along the \aax{} rather than the \cax{}.
	We have shown that this phase transition occurs at uniaxial stress levels which are easily accessible in the laboratory.
	Through thermodynamic and transport measurements, we have established a phase diagram and identified several phase transitions; a second order transition to the \qa-only state upon cooling under uniaxial stress, and a first-order transition between the \qc{} and \qa{} states induced by uniaxial stress for temperatures below \ti{}.
	The observed phenomena map directly to a sixth-order free-energy expansion.
	This work motivates the application of x-ray scattering and ARPES studies of \rte{} under strain to further quantify the behavior and characteristics of the strain-induced \qa{} state.
	Extension of the techniques presented here to samples with quenched disorder, such as Pd\textsubscript{x}\rte{} may also provide unique insight into forms of frustrated charge order on a tetragonal lattice under uniaxial stress.

\section{Acknowledgments}
The authors wish to thank Steve Kivelson for insightful discussions.
This work was supported by the United States Department of Energy, Office of Basic Sciences, under Contract No. DE-AC02-76SF00515.
Part of this work was performed at the Stanford Nano Shared Facilities (SNSF)/Stanford Nanofabrication Facility (SNF), supported by the National Science Foundation under award ECCS-1542152.


%\bibliographystyle{apsrev4-2}
%\bibliography{RTe3_RZB_v10}
%\bibliography{C:/Users/jstra/Desktop/01_Straquadine_ErTe3_RZB_PRX-20201029T081703Z-001/01_Straquadine_ErTe3_RZB_PRX/RTe3_RZB_v5}

\appendix

\appendix

\section{Details of sample preparation}\label{app:methods}
The in-plane $a$- and $c$-axes of samples are distinguished using x-ray diffraction by comparing amplitudes of the (061) peak with its forbidden counterpart (160).
Samples were epoxied and clamped to the stress cell between a set of roughened titanium mounting plates.
The sample is electrically isolated from the bottom titanium mounting plates with small pieces of thin tissue paper impregnated with epoxy.
The distance between the edges of the clamps is approximately \SI{1}{mm}.

We include the effect of thermal contraction of the plates of the capacitive sensor itself, and at any given temperature we use the measured plate spacing as the sample length $L$.

This protocol, while complicated, minimizes the time allowed for thermal and mechanical drifts in the apparatus.
The voltage-strain relationship of the PZT stacks depends strongly on temperature and voltage history, and a smooth slow temperature ramp, rather than a series of temperature steps and the associated equilibration time at each step, produces the widest, highest resolution coverage of the stress-temperature space in the shortest time.
The cases of increasing and decreasing voltages differ by a slight hysteresis ascribed to the presence of a first order transition (to be discussed in detail below) but otherwise show the same qualitative behavior.
\Cref{fig:hyst} shows a direct comparison of data taken on increasing and decreasing voltage steps, and \cref{fig:erte3_ece}(b) shows critical temperatures extracted from both increasing and decreasing voltage steps.

Electrical contacts are formed by sputtering gold onto the freshly cleaved surface, and gold wires are attached with \textit{DuPont} 4929 silver paint.
All of the contacts are placed within the suspended section of the crystal to minimize contributions from the clamped regions, which may experience significant strain inhomogeneity.
The resistivity and elastoresistivity are extracted simultaneously using the demodulation techniques presented in \onlinecite{Hristov2018}.
The voltage signal is amplified through an SR554 transformer from \textit{Stanford Research Systems} with gain of 100.
The transformer's frequency dependence is independently calibrated and measured signals are corrected to reflect this.

%	\paragraph{Elastocalorimetry}
We detect strain-induced oscillations in the sample temperature using a thermocouple.
A Type E thermocouple\cite{Burns1993} is formed by spot-welding pieces of constantan and chromel thermocouple wire, both \SI{12.5}{\micro m} in diameter.
The welded junction is then attached to the center of the top face of the sample using either two-part epoxy or silver paint.
The reference junction is formed by attaching the free ends of the wires to copper pads which are thermalized to the body of the strain cell, as in ref. \onlinecite{Ikeda2019}.
The detected signal is amplified with an SR554 transformer as well as an SR560 preamplifier, together providing a composite gain of 2000.

\section{AC capacitance measurements}\label{app:capbridge}

	\begin{figure}
		\centering
		\includegraphics[width=\columnwidth]{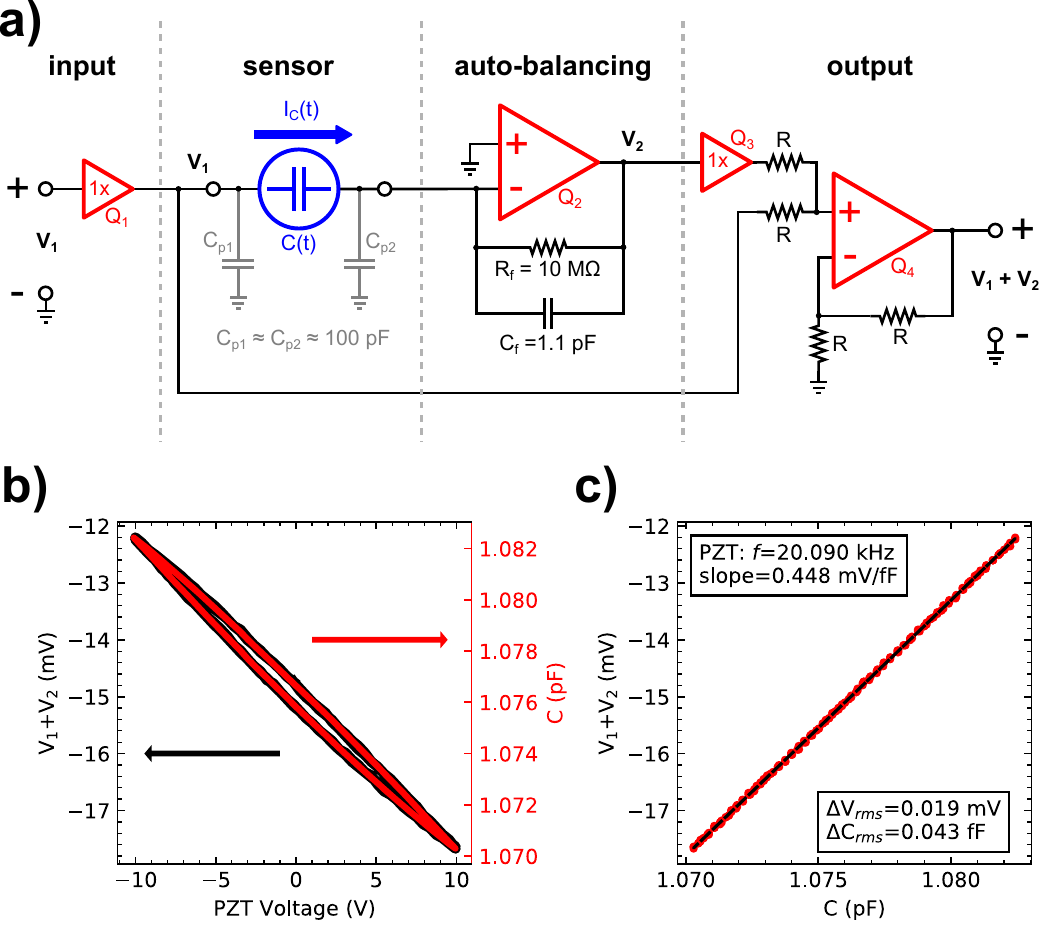}
		\caption{Construction and calibration of the auto-balancing capacitance bridge.
			(a) Diagram of the circuit employed here.
			A $\approx$20~kHz reference input sine wave ($V_1$) is buffered at the input, then applied to one terminal of the displacement sensor through coaxial cables.
			The opposite terminal is connected to a virtual ground by the auto-balancing section, where the feedback capacitor acts as a reference.
			The output $V_2$ of the auto-balancing bridge will be a sine wave with magnitude proportional to the ratio of the sensor and reference capacitances, but $180^{\circ}$ out of phase with the reference.
			The signal at $V_1$ and $V_2$ are added together with a non-inverting summing amplifier, and the two will cancel if the sensor and reference capacitances are equal.
			(b) Output of the capacitance bridge (right axis) overlaid with independent quasistatic measurements made with a commercial capacitance bridge.
			(c) The output voltage plotted against the measured capacitance, demonstrating linear behavior and a sensitivity of \SI{448}{\micro V/fF} and a root-mean-square noise floor of 43~aF.
			This corresponds to a displacement of approximately 2~nm.
		}
		\label{fig:capbridge}
	\end{figure}

	The CS-100 uniaxial strain cell from \textit{Razorbill Instruments} incorporates a capacitive displacement sensor used to quantify the strain in the sample.
	Highly accurate measurements of the capacitance can be made for quasi-static cell displacements, but techniques for measuring changes in capacitance occurring at frequencies above a few Hz are not well established.
	To make it possible to accurately quantify the oscillating strains, and therefore of the elastoresistivity and elastocaloric effect, we have developed a custom bridge circuit (based on a commonly used impedance measurement technique) which, in conjunction with a pair of lock-in amplifiers, measures both DC and AC changes in capacitance.
	The bridge is based on an auto-balancing bridge topology, which can effectively cancel effects of parasitic capacitances due to the cables.

	A simplified circuit schematic is shown in \cref{fig:capbridge}.
	An AC voltage (amplitude 0.5~V, frequency around 20~kHz) is applied to the input terminal, passed through a unity-gain buffer $Q_1$ to minimize the output impedance, and then passed to the CS-100 displacement sensor through a coaxial cable.
	The capacitance of the sensor is of order 1~pF and varies approximately $\pm$~250~fF through the full range of displacements.
	In comparison, typical capacitances for the coaxial cables required to connect the circuit (at room temperature) to the stress cell inside the cryostat are of order 100~pF, more than two orders of magnitude larger than the changes to be observed.
	The low output impedance of the first buffer stage prevents phase lag due to charging and discharging of this parasitic cable capacitance $C_{p1}$.

	The other terminal of the capacitive sensor is connected to a virtual ground at the inverting terminal of the auto-balancing stage.
	The feedback network of this stage contains capacitor which acts as the reference, as well as a resistor for stability.
	The output of this stage is such that the second terminal of the capacitive sensor remains at zero volts, implying that the parasitic capacitance $C_{p2}$ at this terminal does not draw any current.
	The output voltage $V_2$ required for this is proportional to the ratio of the sensor capacitance $C(t)$ to the reference capacitance $C_f$, but with opposite sign.
	The final stage combines the output $V_2$ with the input reference signal such that the two cancel exactly when $C(t)=C_f$.

	The magnitude of the sum is detected by a lock-in amplifier.
	Knowing $C_f$, the time-averaged magnitude can then be converted to provide a measurement of the DC capacitance.
	The instantaneous magnitude as it appears at the output of the lock-in amplifier, however still contains the modulation at the strain frequency; a second demodulation stage of this signal provides direct access to the magnitude of the AC capacitance variation.

	We have calibrated our bridge circuit against a commercial capacitance bridge for quasi-static strain, as shown in \cref{fig:capbridge}(b) and (c), and found that the behavior is well within the linear regime.
	The capacitance-to-voltage conversion factor is approximately 0.448~mV/fF, and the noise floor is approximately 43~aF.
	By simulating the strain-modulated capacitance sensor with an amplitude-modulated current source, we have also confirmed that the circuit described here maintains its accuracy to within 2\% for strain frequencies up to 1~kHz.
	Typical changes in capacitance for a 5~V\rms{} oscillating voltage on the outer pair of piezoelectric stacks range between 0.75~fF{} at 20~K to approximately 2.5~fF\rms{} at room temperature.

\section{Effects of thermal expansion mismatch, finite epoxy stiffness}\label{app:nonideal}
	%	\paragraph{Strain nonideality}
	In an ideal case, where the stress cell and epoxy would be infinitely stiff compared to the sample, the sample strain is given by $\varepsilon=\Delta L/L$, where $L$ is the sample length and $\Delta L$ is the displacement detected by the capacitive sensor.
	This case would be equivalent to the thermodynamic condition of fixing constant strain along the long axis of the sample (temporarily defining this as the $x$-axis $\varepsilon_{xx}$) while the other components of the strain tensor are allowed to relax.
	This differs slightly from the conditions described in the guiding model of the previous section, in which $\tilde{\sigma}_{xx}$ was held constant.
	In a free-standing crystal, the coupling terms between the CDW and strain ($\lambda$ and $\eta$ in \cref{eq:gcoupling}) cause strain to behave like a secondary order parameter.
	A finite value of the CDW gap in either direction induces a sympathetic orthorhombic distortion.
	Fixing constant strain rather than stress would largely preserve the phase diagram of \cref{fig:glrte3}(f) except near the \qa{}-\qc{} transition--the first order transition would widen to encompass a region characterized by a patchwork of orthogonal domains of $|\tilde{\varepsilon}_A|>0$ such that the average strain matched the externally imposed condition.
	In reality, however, we must mention three caveats regarding this idealized constant-strain condition.

	First, the pliability of the epoxy is known to decrease the strain transmitted to the sample.
	Simulations for the case of iron-pnictide superconductor samples show strain transmission of approximately 70\% the ideal value\cite{Ikeda2018}.
	This effect changes quantitative estimates of critical values of the strain to induce the phase transition shown in \cref{fig:glrte3}(f), but does not affect the overall features of the phase diagram.
	For the present study we therefore neglect this effect since our focus is on general features of the strain-tuned phase diagram.

	Secondly, mismatch of the thermal expansion coefficients between the sample and titanium cell body generates a temperature-dependent stress.
	In the case of \rte{}, the in-plane thermal expansion is approximately five times larger than that of titanium at room temperature\cite{Ru2008b}.
	While the thermal expansion of \rte{} has not been measured for all temperatures, it is sensible to assume that the estimated magnitude of tensile (compressive) strain is always underestimated (overestimated).
	In the absence of a direct measure of the \rte{} lattice parameter, we have chosen not to apply any offset corrections to the reported strains throughout this work.
	Doing so would require extrapolations from diffraction measurements on freestanding samples, but such extrapolations would not apply to a state in which the CDW has been rotated by strain.
	The cost of neglecting the sample thermal expansion in this calculation is a negligible underestimate of the magnitude of the strain oscillations, and the inclusion of an unknown and temperature dependent strain offset.

	Finally, the process of curing the mounting epoxy can produce built-in strains even at room temperature, and this offset varies from sample to sample.
	In order to compare samples on the same scale, we extract the location of the critical point between the \qc{}, \qa{}, and disordered states, which we define as occurring at the degeneracy strain $\varepsilon_{aa}^0$.
	The arguments provided in this paper do not require the exact knowledge of the absolute strain, but rather focus on the relative changes.

\section{Hysteresis measurements in second sample} \label{sec:hyst2}

	\begin{figure}
		\centering
		\includegraphics[width=\columnwidth]{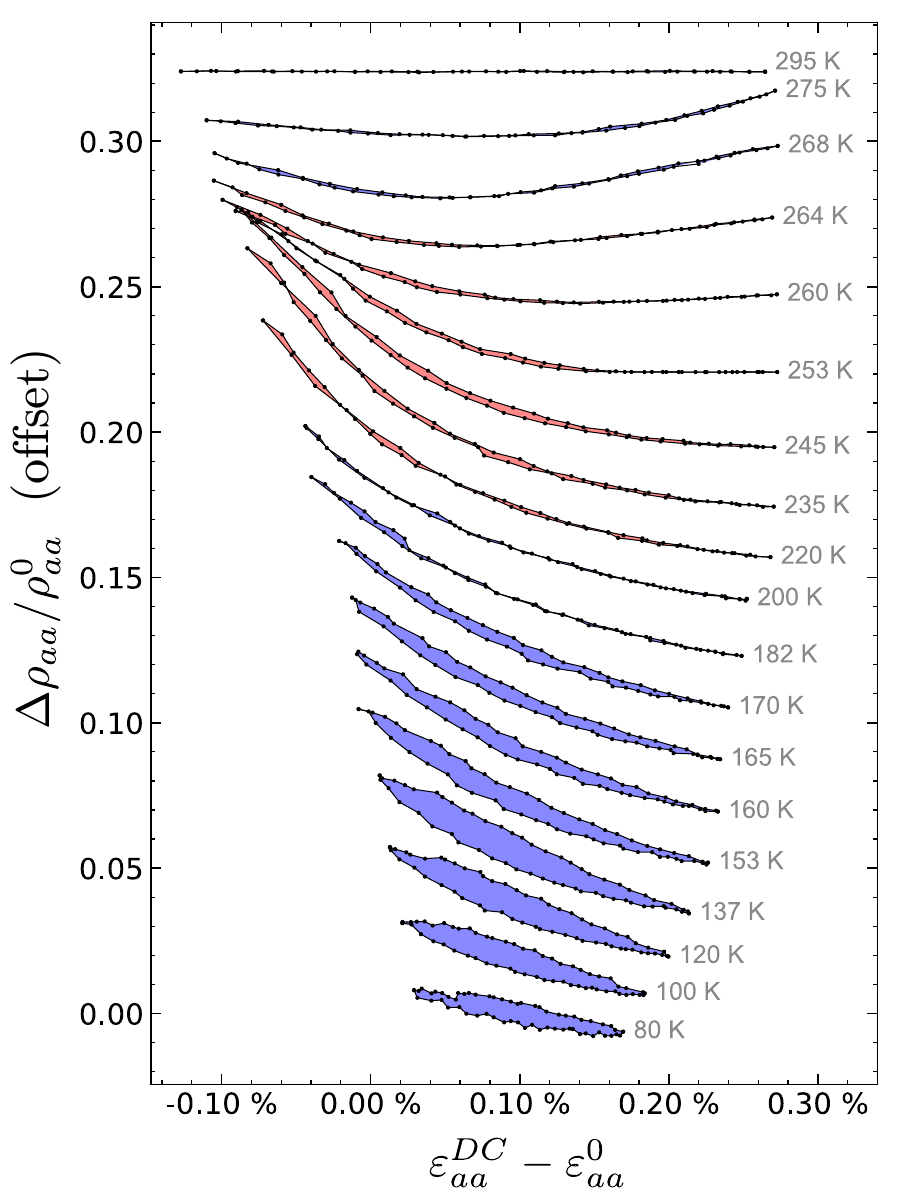}
		\caption{
			Resistivity stress-cycle data for a second sample of \er{}, showing a similar phenomenon, but with a decreased magnitude of the effect.
			We attribute this difference to differences in the accessible range of strains in our experimental setup.
		}
		\label{fig:hyst2}
	\end{figure}

		\Cref{fig:hyst2} shows the stress-cycle data for the \aax{} resistivity of a second \er{} sample, similar to \cref{fig:hyst}.
		As seen in the previous sample, hysteresis becomes apparent below \ti{}, then changes directions on further decreases of temperature.
		In this sample, due to subtle differences in experimental details (see \cref{app:nonideal}) the accessible range of strain does not appear to include the lower coercive strain; the resistivity is not seen to saturate on the compressive side as in \cref{fig:hyst}.
		As a result, the magnitude of the hysteresis is substantially decreased.
		The qualitative structure, however, appears largely the same.

\section{Frequency dependence of ECE measurements}\label{sec:freqs}
	\begin{figure}
		\centering
		\includegraphics[width=\columnwidth]{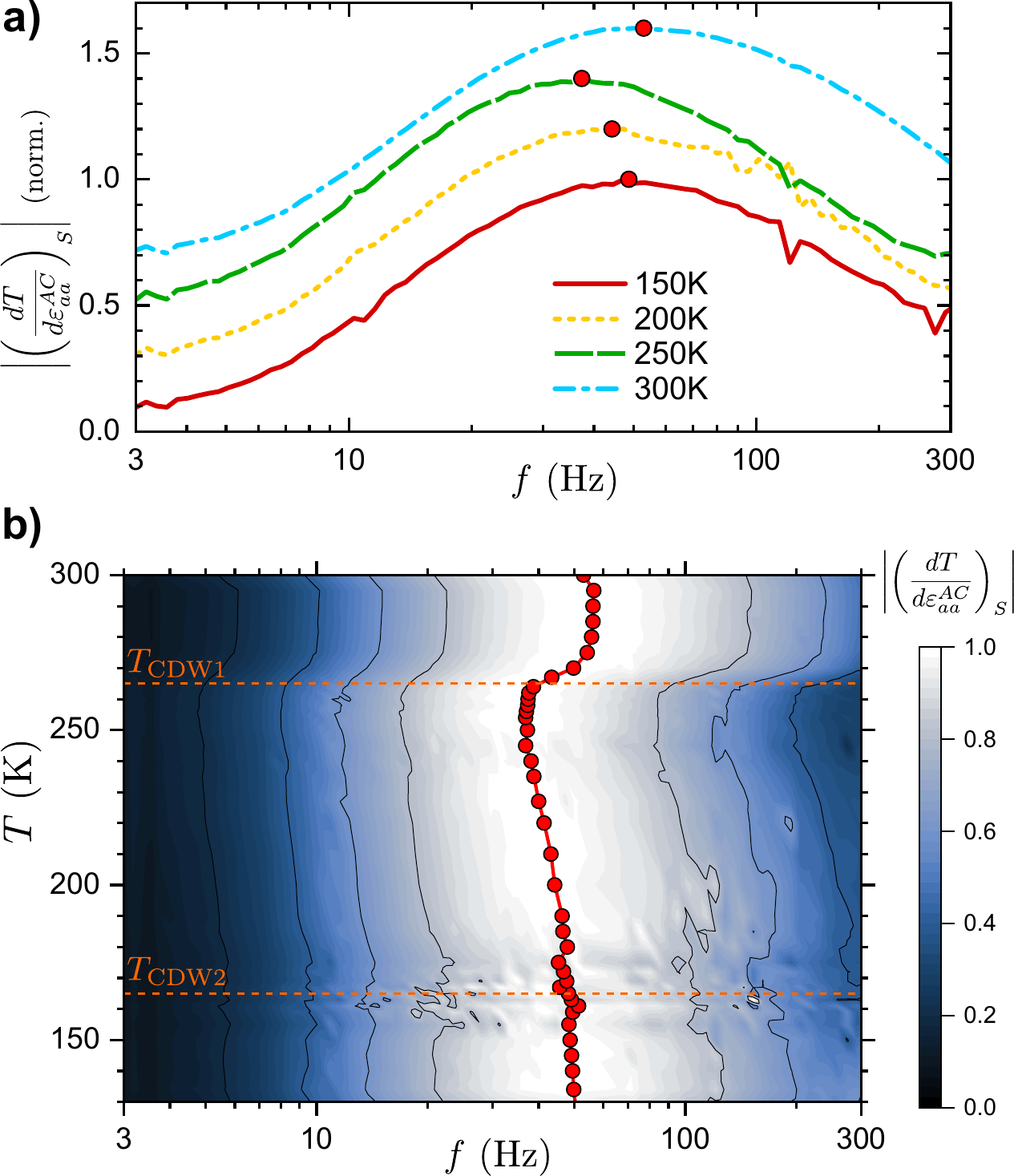}
		\caption{
			Frequency dependence of elastocaloric effect measurements on \er{} under \aax{} stress.
			(a) Magnitude of the ECE signal as a function of strain frequency for several representative temperatures.
			Each curve is normalized to the maximum signal observed at that temperature, and the curves are vertically offset for clarity.
			The peak is notated with a red circle.
			Panel (b) shows similar data for a dense set of temperatures, presented as a contour plot.
			As the sample cools below \ti{}, the peak frequency decreases from approximately 55~Hz to approximately 35~Hz.
			This shift in the peak frequency reflects a sudden decrease in thermal conductivity.
			This is qualitatively consistent with the Wiedemann-Franz law: this measurement is most sensitive to the thermal conductivity along the \aax, and the electrical resistivity $\rho_{aa}$ also shows the highest increase when cooling through \ti.
			Data presented in the main text was acquired for a constant frequency in the range of $20-40$~Hz.
			Shifts in the frequency dependence with temperature will affect the absolute magnitudes of the ECE signals in \cref{fig:erte3_ece,fig:tmte3_ece}, but would alter neither qualitative behavior nor the conclusions stemming from this data.
		}
		\label{fig:freqs}
	\end{figure}

	As described in detail in ref. \onlinecite{Ikeda2019}, the criteria for quasi-adiabatic behavior in ECE measurements depend on the thermal properties of the sample, as well as the properties of the materials used to mount the sample and to adhere the thermocouple.
	At low strain frequencies, the sample tends to thermalize with the mounting plates, decreasing the observed temperature oscillation magnitude.
	At high strain frequencies, the thermocouple can no longer follow the changes in temperature.
	The maximum signal is observed at intermediate frequencies.
	Formally, the low frequency cutoff for quasi-adiabatic behavior $\omega_{qa}$ depends on the thermal conductivity $\kappa$, the volumetric specific heat $\rho c_p$, and the sample length $L$ as
	\begin{equation}\label{eq:omegaqa}
		\omega_{qa} \propto \frac{\kappa}{\rho c_p L^2}.
	\end{equation}
	As a consequence, any changes in the thermal conductivity or specific heat will result in a change of the observed elastocaloric signal.

	Upon cooling, as one passes through the CDW transition temperature \ti{}, the electrical resistivity is shown in \cref{fig:rhoinplane} to change significantly, especially $\rho_{aa}$.
	The Wiedemann-Franz law suggests that a similar change would be observed in the thermal conductivity.
	The temperature dependence data presented in \cref{fig:erte3_ece,fig:tmte3_ece} was taken using a single frequency for the entire temperature and strain range; the temperature dependence of the thermal conductivity should be expected to alter the signal.

	\Cref{fig:freqs} shows the frequency dependence of magnitude of the elastocaloric effect for \aax{} stress on \er{} as a function of temperature.
	The frequency $f_p$ at which the ECE reaches its peak is denoted by the filled symbols.
	As the sample is cooled through \ti{}, the location of the maximum moves to lower frequencies, corresponding to a change in the ratio $\kappa/c_p$.
	The heat capacity, which is dominated by the phonon background (the Debye temperature $\Theta_D\approx 180$~K\cite{Ru2006,Banerjee2013})) is known to be approximately constant through \ti{}\cite{Saint-Paul2017}.
	Therefore the change in the frequency dependence must arise from changes in the thermal conductivity.

	Two different contributions to the ECE anomaly at \ti{} were discussed in the main text: critical fluctuations near a strain-tuned phase transition, and changes in the thermal expansion tensor.
	For an ECE measurement at a constant frequency $f_0$, however, a sharp change in the frequency dependence can also contribute to this anomaly.
	If $f_0<f_p$, the frequency dependence would cause the ECE magnitude to rise upon cooling through \ti{}.
	The ECE measurements in \cref{fig:erte3_ece,fig:tmte3_ece} were made with strain frequencies of 37~Hz and 51~Hz, respectively.
	The ECE anomaly in the near-freestanding case does indeed increase the ECE magnitude on cooling through \ti{}, implying that a frequency dependence could indeed be contributing to the behavior.
	However, both the \qc{} and strain-induced \qa{} states should correspond to a decrease in thermal conductivity due to the opening of a CDW gap.
	Therefore we conclude that changes in the frequency dependence can decrease the size of the ECE anomaly, but such changes cannot explain the change in sign of the anomaly.
	Frequency dependent effects can at most alter the absolute magnitude, but the effects discussed in the main text must still be the primary contributions to the ECE.

\end{document}